\documentclass[prd,aps,nofootinbib]{revtex4}
\usepackage{graphicx}
\begin{document}

\title{Trans-Planckian Dark Energy?}

\author{Martin Lemoine}

\affiliation{Institut d'Astrophysique de Paris, GReCO, CNRS-FRE 2435,
\\ 98 bis Boulevard Arago, F-75014 Paris, France}

\author{J\'er\^ome Martin} 

\affiliation{Institut d'Astrophysique de Paris, GReCO, CNRS-FRE 2435,
\\ 98 bis Boulevard Arago, F-75014 Paris, France}

\author{Jean-Philippe Uzan}

\affiliation{Institut d'Astrophysique de Paris, GReCO, CNRS-FRE 2435 
\\ 98 bis Boulevard Arago, F-75014 Paris, France
\\ and \\
Laboratoire de Physique Th\'eorique, CNRS-UMR 8627, 
Universit\'e Paris Sud,
\\ B\^atiment 210, 91405 Orsay Cedex, France.}

\date{\today}
\begin{abstract}

It has recently been proposed in Refs.~\cite{MBK01,BFM02,BM02} that
the dark energy could be attributed to the cosmological properties of
a scalar field with a non-standard dispersion relation that decreases
exponentially at wave-numbers larger than Planck scale ($k_{\rm
phys}>M_{\rm Pl}$). In this scenario, the energy density stored in the
modes of trans-Planckian wave-numbers but sub-Hubble frequencies
produced by amplification of the vacuum quantum fluctuations would
account naturally for the dark energy. The present article examines
this model in detail and shows step by step that it does not work. In
particular, we show that this model cannot make definite predictions
since there is no well-defined vacuum state in the region of
wave-numbers considered, hence the initial data cannot be specified
unambiguously. We also show that for most choices of initial data this
scenario implies the production of a large amount of energy density
(of order $\sim M_{\rm Pl}^4$) for modes with momenta $\sim M_{{\rm
Pl}}$, far in excess of the background energy density. We evaluate the
amount of fine-tuning in the initial data necessary to avoid this
back-reaction problem and find it is of order $H/M_{{\rm Pl}}$. We
also argue that the equation of state of the trans-Planckian modes is
not vacuum-like. Therefore this model does not provide a suitable
explanation for the dark energy.
\end{abstract}
\pacs{} \maketitle

\section{Introduction}\label{Sec:I}

Recently, it has been claimed in a series of
articles~\cite{MBK01,BFM02,BM02} that the cosmic dark energy component
could be explained naturally by the trans-Planckian energy of a scalar
field with a suitable non-linear dispersion relation in the
trans-Planckian regime. Such dispersion relations, which relate the
frequency $\omega_{\rm phys}$ to the wave-number $k_{\rm phys}$ of a
scalar field wave-packet, and which depart from the standard linear
dispersion relation in the trans-Planckian regime are a way of
modeling phenomenologically the unknown physics for sub-Planckian
wavelengths. They have been used extensively in the recent literature
in the context of black-hole physics~\cite{U95} and of the
inflationary trans-Planckian problem~\cite{MB}.

In the particular case considered in Refs.~\cite{MBK01,BFM02,BM02},
the dispersion relation departs from its standard linear form and
approach a decreasing exponential at large wave-numbers. This type of
dispersion relation could possibly emerge from string
theory~\cite{BFM02}. It has been argued that the energy density of the
modes of sub-Planckian wavelengths and sub-Hubble frequencies
(referred to as ``tail'' modes) is naturally of the same order as the
critical energy density today and has the same equation of state as a
cosmological constant.  Hence, it could account without fine-tuning
for the dark energy.  The energy density contained in the tail today
has been calculated in Ref.~\cite{MBK01} by solving for the time
evolution of a test quantum scalar field evolving in the curved
cosmological background, assuming its initial state at the onset of
inflation is the vacuum. The equation of state of the tail modes has
been calculated in Ref.~\cite{BM02} and its cosmological evolution has
been solved to argue that the cosmic coincidence problem (why does the
dark energy dominate now?) is solved.

In this paper we argue that this model does not and can not work for
several reasons. We first argue that the energy density and equation
of state of the ``tail'' modes depend directly on the choice of
initial data for the scalar field, and that this latter cannot be
specified {\it unambiguously} since there is no preferred initial
vacuum state (Section~\ref{Sec:II}). This implies that any
cosmological consequence derived depends directly on the ad-hoc choice
of initial data (initial quantum state). We then show that the
violation of the Wentzel-Kramers-Brillouin (WKB) approximation in the
remote past for all co-moving wave-numbers, which is inevitable in the
present scenario, implies the continuous production {\it at all times}
of a large amount of quanta with physical wave-number $\sim M_{{\rm
Pl}}$ (Section~\ref{Sec:III}). This finding is in agreement with
general arguments given by Starobinsky~\cite{S01} (this latter work
did not however study the present scenario). We evaluate the amount of
energy density produced for modes of physical wave-number and
frequency $\sim M_{{\rm Pl}}$ for various choices of initial data and
conclude that it is generically of order $M_{{\rm Pl}}^4$. This
process takes place at all times, and since the energy density
produced is much larger than the background energy density, it implies
that the semi-classical perturbative framework on which the model of
Refs.~\cite{MBK01,BFM02,BM02} rests breaks down.  In an earlier study,
devoted to constructing an effective stress-energy tensor for theories
with non-linear dispersion relations~\cite{LMMU01}, we already
criticized this model by arguing that it led generically to the wrong
equation of state. We revisit this issue further in
Section~\ref{Sec:IV}, where we prove that the effective
energy-momentum tensor we derived earlier is well-behaved, thus
disproving an improper claim of Ref.~\cite{BM02} and confirming our
earlier criticisms. We provide a summary of our conclusions in
Section~\ref{Sec:V}.

\section{Initial conditions for the mode evolution}\label{Sec:II}

\begin{figure}
      \centering
      \includegraphics[width=0.7\textwidth,clip=true]{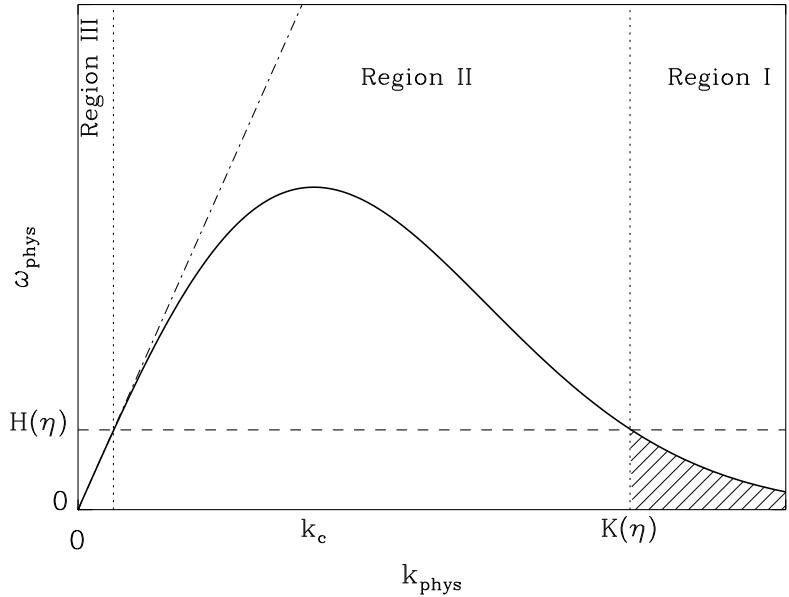}
      \caption[...]{Dispersion relation $\omega_{\rm phys}$ {\it vs}
      $k_{\rm phys}$. Region I (hashed area) corresponds to the
      ``tail'' modes for which $\omega_{\rm phys}\leq H$ [and $k_{\rm
      phys}\geq K(\eta)$]. Region II corresponds to the sub-horizon modes
      ($k_{\rm phys}\geq H$) that are outside of the tail, while
      region III corresponds to the super-horizon modes $k_{\rm
      phys}\leq H$. The fundamental scale $k_{\rm c}$ is also indicated.}
\label{fig0}
\end{figure}

Refs.~\cite{MBK01,BFM02,BM02} consider a scalar field, $\phi$, with a
non-linear dispersion relation that is linear in the sub-Planckian
regime and approaches a decreasing exponential at trans-Planckian
wave-numbers (i.e. for $k_{\rm phys}\gtrsim k_{\rm c}$, $k_{\rm c}
\sim M_{{\rm Pl}}$ being a fundamental characteristic scale). This
dispersion relation is shown in Fig.~\ref{fig0}.  This scalar field is
assumed to describe the density (scalar) perturbations and/or the
primordial gravitational waves. The ``tail'' modes are thus
interpreted as a bath of gravitons of super-Planckian wavelengths and
sub-Hubble frequencies. This scalar field is treated as a test-field
(its back-reaction on the background is neglected) and is quantized on
the curved cosmological background. Assuming that the ``tail'' modes
of this field are initially in a well chosen vacuum state as
$\eta\rightarrow -\infty$ ($\eta$ denoting conformal time), the
occupation number at late times ($\eta\rightarrow+\infty$) of quanta
extracted out of the vacuum by the dynamical background has been
calculated in Ref.~\cite{MBK01}. This occupation number can then be
used to calculate the energy density stored today in the
``tail''. This is the thread of the calculation performed in
Ref.~\cite{MBK01}, which we now follow in some detail. This discussion
will take us to the two main arguments that we bring forward against
this model (given in this Section and the following).

The equation of motion of a scalar field $\phi\equiv \mu /a $ in a
Friedmann-Lema\^{\i}tre-Robertson-Walker (FLRW) space-time with scale
factor $a(\eta)$ reads:
\begin{equation}\label{eq:1}
 \mu_k'' + \left[\omega^2 - (1-6\xi)\frac{a''}{a}\right]\mu_k=0,
\end{equation}
where $\xi$ is a coupling parameter to gravity and a prime denotes
differentiation with respect to conformal time. $k=ak_{\rm phys}$ is
the co-moving wave-number and $\omega=a\omega_{\rm phys}$ is the
co-moving frequency. $\xi=0$ for tensor perturbations degrees of
freedom and $\xi=1/6$ for a conformally coupled field. The scale
factor is taken to be a power law in conformal time,
$a=\vert\eta/\eta_{\rm c}\vert^{-\beta}$, and the following dispersion
relation, parameterized by two parameters $\epsilon_1$ and
$\epsilon_3$ with $\epsilon_3=4-2\epsilon_1$ in order to insure that
the dispersion relation is linear for small wave-numbers, is
introduced:
\begin{equation}
\omega_{\rm phys}^2(k_{\rm phys})=k_{\rm
phys}^2\left[\frac{\epsilon_1}{1+{\rm e}^x} +
\frac{\epsilon_3 {\rm e}^x}{(1+{\rm e}^{x})^2}\right],
\end{equation}
with $x\equiv (k_{\rm phys}/k_{\rm c})^{1/\beta}=A\vert\eta\vert$ with
$A\equiv (k/k_{\rm c})^{1/\beta}/\vert \eta_{\rm c}\vert$. A problem
with this dispersion relation is that it depends on the power-law
index $\beta$ of the scale factor. If taken literally, this means that
the dispersion relation, or the physical frequency of a given mode,
changes as the scale factor power-law index $\beta$ changes between
various cosmological eras (e.g. inflation / radiation domination /
matter domination). More importantly one easily sees that the above
dispersion relation has a pathological behavior in the radiation
($\beta=-1$) or matter ($\beta=-2$) dominated eras.  In fact, for
$\beta<0$, it implies $\omega_{\rm phys}/k_{\rm phys}\rightarrow 0$ as
$k_{\rm phys}\rightarrow 0$, whereas one should instead reach the
linear dispersion relation in that regime with $\omega_{\rm
phys}/k_{\rm phys}\rightarrow1$. Since Ref.~\cite{MBK01} focused on
the case of de Sitter space-time with $\beta=1$, we set $\beta=1$ in
the above dispersion relation, i.e., the above $x$ should be
understood as $x\equiv k_{\rm phys}/k_{\rm c}$.  This reformulated
dispersion relation thus coincides with that used in Ref.~\cite{MBK01}
for de Sitter space. However, in the matter dominated era, for
instance, we have $x=k_{\rm phys}/k_{\rm c}\propto \eta^{-2}$ and the
general class of solutions to the field equation obtained in
Ref.~\cite{MBK01} does not hold anymore. The linear dependence of $x$
on $\eta$ is lost for background metrics other than de Sitter, but the
linear dependence of $\omega_{\rm phys}$ on $k_{\rm phys}$ is
preserved in the small $k_{\rm phys}$ limit for all metrics, which is
obviously an imperative. The field equation can finally be rewritten
as:
\begin{equation}\label{eq:2}
 \mu _k''+\biggl\{k^2\biggl[\frac{\epsilon _1} {1+{\rm e}^{x}} +
\frac{\epsilon_3{\rm e}^{x}}{ (1+{\rm e}^{x})^2}\biggr]
-(1-6\xi )\frac{\beta(\beta+1)}{\eta ^2}\biggr\}\mu _k=0\, ,
\end{equation}
with, again, $x(\eta)\equiv k_{\rm phys}/k_{\rm c}=k/[a(\eta)k_{\rm
c}]$. The solution to this equation depends on the value of $\xi$ and
$\beta$. In Ref.~\cite{MBK01}, the contribution of the $[a''/a]$ term
is assumed to be negligible at early times. However the above equation
shows that this is not the case; denoting by $\Omega_k^2$ the term in
curly brackets in Eq.~(\ref{eq:2}), one has $\Omega_k^2\simeq
k^2(\epsilon_1+\epsilon_3){\rm e}^{-k\vert\eta\vert^\beta/(k_{\rm
c}\vert\eta_{\rm c}\vert^\beta )} -\beta(\beta+1)(1-6\xi )/\eta ^2$ as
$\eta \rightarrow -\infty $ and the term $a''/a\propto \eta ^{-2}$ is
always dominant in that limit if $\xi\neq1/6$ ($\beta\geq1$,
$\eta\rightarrow-\infty$). Therefore, in the limit $\xi\neq1/6$ and
$\eta\rightarrow -\infty$ the two independent solutions to the field
equation are power-laws in $\eta$:
\begin{equation}\label{eq:power-law}
 \mu_k \propto \left(\frac{\eta}{\eta_{\rm c}}\right)^{\alpha{_{\pm}}},\quad
 \alpha_\pm = \frac{1}{2}\pm \sqrt{\frac{1}{4} + \beta(1+\beta)(1-6\xi)}\, .
\end{equation}
This is an important point as it implies that the mode function does not
behave as a plane wave in the limit $\eta\rightarrow -\infty$ when
$\xi\neq1/6$.  The solution to the field equation in this limit is
reminiscent of the mode freezing in inflationary theories for fields
with linear dispersion relations and $\xi=0$ when the mode exits the
horizon.

It is also argued that the term in $a''/a$ can be absorbed at late
times in a redefinition of the dispersion relation. However this
cannot be correct since by construction, the dispersion relation
$\omega_{\rm phys}(k_{\rm phys})$ depends only on $k_{\rm phys}$,
i.e. time only enters via $k_{\rm phys}$. Therefore, one can absorb a
term $a''/a\propto \eta^{-2}$ into $\omega_{\rm phys}^2/a^2$ only if
$\beta=1$ (de Sitter space-time), as inspection of Eq.~(\ref{eq:2})
reveals. In effect, the curly bracket of Eq.~(\ref{eq:2}) can then be
rewritten as $a^{-2}$ times a function of $k\eta\propto k_{\rm
phys}$. But, in that particular case, the redefined modified
dispersion relation does not have an exponential shape anymore, since
$\omega_{\rm phys}$ approaches a constant ($\sim H$) as $k_{\rm
phys}\rightarrow + \infty$. However, this should not give the
impression that the corresponding solution is a plane wave since,
evidently, the co-moving frequency $\omega $ which enters
Eq.~(\ref{eq:1}) still behaves as $\propto \eta ^{-2}$. Moreover in
the case of a matter or radiation dominated cosmology, one cannot
absorb the scale factor term in the dispersion relation.

\par

Nevertheless one can also assume $\xi =1/6$. In that case, it is
possible to find an exact solution to the equation of motion in de
Sitter space-time. Indeed, for a conformally coupled field, the term
$a''/a$ disappears from the field equation and the equation becomes
simpler. Note however that the scalar field cannot correspond to
tensor perturbations degrees of freedom since these are minimally
coupled to the metric. Let us consider $\xi=1/6$ for the moment. A
solution to the field equation, given in Ref.~\cite{MBK01}
reads\footnote{Equation~(25) in Ref.~\cite{MBK01} contains a misprint
that has been corrected in the following equation}:

\begin{equation}\label{eq:4}
 \mu_k^{\rm (in)}(\eta) = C^{(\rm in)}
 \left(1+{\rm e}^{-x}\right)^{d+{1\over 2}} {}_2F_1\left(b+d+{1\over
 2},-b+d+{1\over 2};2d+1;1+{\rm e}^{-x}\right)
\end{equation}
where ${}_2F_1$ is an hypergeometric function and $b$ and $d$ are
expressed in terms of $\epsilon_1$ and $\epsilon_3$ as:

\begin{equation}
b\,\equiv\, i \sqrt{\hat\epsilon_1},\quad
d\equiv\sqrt{\frac{1}{4}+\hat\epsilon_3},\quad{\rm and}\quad
\hat\epsilon_i \equiv k_{\rm c}^2\eta_{\rm c}^2\epsilon_i\,.
\end{equation}
This solution is valid only for de Sitter space-time with
$x=k\eta/k_{\rm c}\eta_{\rm c}$ ($\eta<0,\,\eta_{\rm c}<0$). As
already mentioned, this is due to the fact that, with the reformulated
dispersion introduced above, the linear dependence of $x$ in the
conformal time is lost for other scale factors. However similar
solutions for other metrics can be obtained if the dispersion relation
is tuned to the power-law evolution of the scale factor, i.e. if the
parameter $x$ remains linear in $\eta$ (possibly at the expense of
linearity of $\omega_{\rm phys}$ in the small $k_{\rm phys}$ limit,
see above). Equation~(\ref{eq:2}) has in fact two independent
solutions (see below) and the choice~(\ref{eq:4}) represents only one
branch of the solution, which is moreover written on the branch cut of
the hypergeometric function ${}_2F_1$. At early times
($\eta\rightarrow-\infty$, i.e. $x\rightarrow +\infty$), this solution
(\ref{eq:4}) does not oscillate and it blows up\footnote{A
hypergeometric function of the form
$_2F_1(\alpha,\beta;\alpha+\beta;z)$ is singular at $z=1$. One can
also solve Eq.~(\ref{eq:2}) for $\xi =1/6$ and $\beta=1$ in the limit
$\eta \rightarrow -\infty $. In this case, the equation reduces to
$\mu _k''+k^2(\epsilon_1+\epsilon_3){\rm e}^{-x}\mu _k=0$ and the
solution can be written as
\begin{equation}
\label{footeq}
\mu _k(\eta )=\frac{1}{\sqrt{2k}}\biggl[A_1(k)J_0({\rm e}^y)
+A_2(k)N_0({\rm e}^y)\biggr]\, ,
\end{equation}
where $J_0$ and $N_0$ are Bessel functions and where $y\equiv -x/2+\ln
[4k^2(\epsilon_1+\epsilon_3)/A^2]/2$. The Neumann function diverges in
the limit $\eta \rightarrow -\infty$ ($y\rightarrow -\infty$). In the
tail, the corresponding behavior for the scalar field itself is given
by $\phi \propto \eta $ and $\phi \propto \eta ^2 $.}.

It is more convenient to write the general solution to the field
equation, with $\xi=1/6$, as $\mu_k = C_1 \mu^{(1)}_k + C_2
\mu^{(2)}_k$, where $\mu^{(1)}_k$ and $\mu^{(2)}_k$ are two
independent solutions given by:

\begin{eqnarray}\label{eq:mu1mu2}
\mu^{(1)}_k(\eta) & = & {\rm e}^{bx} (1+{\rm e}^x)^{d+1/2}\,
{}_2F_1\left(b+d+\frac{1}{2},b+d+\frac{1}{2};2b+1;-{\rm e}^x\right)\nonumber\\
\mu^{(2)}_k(\eta) & = & {\rm e}^{-bx} (1+{\rm e}^x)^{d+1/2}\,
{}_2F_1\left(-b+d+\frac{1}{2},-b+d+\frac{1}{2};-2b+1;-{\rm e}^x\right).
\end{eqnarray}
Since $b$ is pure imaginary, and $d$ is real, one concludes easily
that $\mu^{(2)}_k=\mu^{(1)*}_k$. The Wronskian of these two solutions
is non-zero, and can be used to relate the coefficients $C_1$ and
$C_2$ so as to obtain canonical commutation relations for the field
operator and its adjoint. Since only one branch of the solution was
given in Ref.~\cite{MBK01}, the canonical commutation relations for
the field and its adjoint could not be satisfied. More precisely, it
can be checked that the solution given in Eq.~(\ref{eq:4}) is real.
This is due to the fact that it involves an hypergeometric function of
the form ${}_2F_1(\alpha ,\alpha ^*; \alpha +\alpha ^*; z)$ with
$z=z^*=1+{\rm e}^{-x}$ and $\alpha\equiv b+d+1/2$ in that case and
${}_2F_1^*(a,b;c;z)={}_2F_1^*(b,a;c;z)={}_2F_1(a^*,b^*;c^*;z^*)$. Therefore,
one has ${}_2F_1^*(\alpha ,\alpha ^*; \alpha +\alpha ^*; 1+{\rm
e}^{-x})={}_2F_1(\alpha ,\alpha ^*; \alpha +\alpha ^*; 1+{\rm
e}^{-x})$ and the mode function is indeed real. It follows that the
Wronskian of the solution considered in Ref.~\cite{MBK01} vanishes:
$W(\mu ,\mu ^*)\equiv \mu _k\mu _k^{*}{}'-\mu _k'\mu _k^*=0$. Using
the properties of hypergeometric functions, one can check that both
independent solutions $\mu^{(1)}_k$ and $\mu^{(2)}_k$ behave as
$\propto x\propto \vert\eta\vert$ in the limit
$\eta\rightarrow-\infty$, i.e. these mode functions blow up. This
result is consistent with Eq.~(\ref{footeq}) since, in the tail, the
two branches $\mu^{(1)}_k$ and $\mu^{(2)}_k$ are linear combinations
of the Bessel functions $J_0$ and $N_0$.

Therefore, we have shown that neither in the case $\xi \neq 1/6$ nor
in the case $\xi =1/6$ does the mode function behave as a plane wave
in the tail. Thus the initial state of the field cannot reduce to the
Bunch-Davies adiabatic vacuum, contrary to the claim~\cite{MBK01}:
``{\it we show that there is no ambiguity in the correct choice of the
initial vacuum state. The only initial vacuum is the adiabatic vacuum
obtained by the solution to the mode equation}''. The usual
prescription to remove the ambiguity on the choice of vacuum state in
curved space-time, i.e.  for constructing a vacuum state which is
closest to the definition of vacuum in Minkowski, is indeed to rely on
the WKB approximation to construct vacua of successively higher
adiabatic order~\cite{BD}.  In this scenario~\cite{MBK01,BFM02,BM02},
this construction cannot be performed for a simple reason: the WKB
approximation, which quantifies the adiabaticity of the quantum mode
evolution is violated at all times for modes contained in the tail,
i.e. modes with $k_{\rm phys}> k_{\rm c}$ and $\omega_{\rm phys}<H$.
The WKB condition can be written in the form $\vert Q/\Omega_k^2 \vert
\ll 1$~\cite{BD}, where $\Omega_k^2$ denotes the term in curly
brackets in Eq.~(\ref{eq:2}) as before, and $Q\equiv
\Omega_k''/2\Omega_k - (3/4)\Omega_k^{\prime 2}/\Omega_k^2$. In effect
the WKB solution $\mu_{\rm WKB}\equiv \exp(\pm i\int \Omega_k {\rm
d}\tau)/\sqrt{2\Omega_k}$ exactly verifies the following equation
$\mu_{\rm WKB}'' + (\Omega_k^2 - Q)\mu_{\rm WKB}=0$. Therefore it is a
good approximation to the solution of the actual mode equation $\mu''
+ \Omega_k^2\mu=0$ if the above inequality $\vert Q/\Omega_k^2 \vert
\ll 1$ is satisfied (see also Ref.~\cite{MS02} for more details).

The expression for $Q/\Omega_k^2$ is cumbersome, but since we are
interested in the regime $k_{\rm phys}\gg k_{\rm c}$, we may use the
limiting form of the dispersion relation:

\begin{equation}\label{eq:omega-approx}
 \omega_{\rm phys}\,\simeq \,k_{\rm
phys}\sqrt{\epsilon_1+\epsilon_3}{\rm e}^{-k_{\rm phys}/(2k_{\rm c})}
\quad\quad (k_{\rm phys}\gg k_{\rm c}),
\end{equation} 
If $\xi=1/6$, then $\Omega_k=a\omega_{\rm phys}$ and $\vert
Q/\Omega_k^2\vert \sim k_{\rm phys}^2H^2/16k_{\rm c}^2\omega_{\rm
phys}^2$. In order to understand the behavior of $\vert
Q/\Omega_k^2\vert$, it is convenient to introduce the physical
wave-number $K_+>k_{\rm c}$ such that $\omega_{\rm
phys}(K_+)=\sqrt{(1+\beta)/\beta}H$ [in the case $\xi=0$, one has
$\Omega_k(K_+)=0$]. This wavenumber $K_+\approx K$, where $K(\eta)$ is
the physical wavenumber that gives the lower limit of the tail,
as indicated in Fig.~\ref{fig0}.  Using Eq.~(\ref{eq:omega-approx}),
one easily derives:

\begin{equation}
K_+\,\simeq\, 2k_{\rm
c}\ln\left[\sqrt{\frac{(\epsilon_1+\epsilon_3)\beta}{1+\beta}}
\frac{2k_{\rm c}}{H}\right].
\end{equation}
This formula is written to zeroth order in $\ln(k_{\rm c}/H)/(k_{\rm
c}/H)$ but can be expanded to arbitrary order in a straightforward
way.  The meaning of the physical wave-number $K_+$ is the following
(see Fig.~\ref{fig0}). If $k_{\rm phys}\ll K_+$ but $k_{\rm phys}>
k_{\rm c}$ (i.e., within region II), the mode is outside of the tail
with $\omega_{\rm phys}\gg H$. If however $k_{\rm phys}\gg K_+$, the
mode is in the tail with $\omega_{\rm phys}\ll H$ (region I of
Fig.~\ref{fig0}). Then $k_{\rm phys}\gg K_+$ means that $\omega_{\rm
phys}^2\ll H^2 $ which implies in turn $\vert Q/\Omega_k^2\vert \gg
k_{\rm phys}^2/(16k_{\rm c}^2) \gg 1$, hence the WKB approximation is
violated at all times in the tail. Note that outside of the tail, i.e.
for $k_{\rm phys}\ll K_+$ and $k_{\rm phys}\gg k_{\rm c}$ (region II
of Fig.~\ref{fig0}), the WKB approximation becomes valid. In effect
$k_{\rm phys}/\omega_{\rm phys}\propto \exp(k_{\rm phys}/2k_{\rm c})
\ll \ln(k_{\rm c}/H)$, hence $\vert Q/\Omega_k^2 \vert \ll 1$.

\par

If $\xi \neq 1/6$, then for $\omega_{\rm phys}\ll H$ (region I or tail
in Fig.~\ref{fig0}), the dominant term is $\propto a''/a$ in the
expression of $\Omega _k^2$, namely $\Omega_k^2\sim -(1-6\xi)(1+\beta
)a^2H^2/\beta $, hence $\vert Q/\Omega_k^2 \vert \sim [4(1-6\xi )\beta
(\beta +1)]^{-1}$, which for $\beta=1$ (de Sitter) and $\xi=0 $
(minimal coupling) reduces to $1/8$. In this case, it can be shown
that the WKB approximation does not give the right behavior for the
mode function even though $\vert Q/\Omega_k^2\vert$ is smaller than
unity~\cite{MS02}, and that the WKB approximation is not valid
either. Again, note that outside of the tail (region II of
Fig.~\ref{fig0}) the WKB approximation is valid. The calculation is
the same as in the previous paragraph, since for $k_{\rm phys}\ll K_+$
and $k_{\rm phys}\gg k_{\rm c}$, one has $\Omega_k^2 \sim
a^2\omega_{\rm phys}^2$ since $\omega_{\rm phys}\gg H$. Thus one finds
$\vert Q/\Omega_k^2\vert\ll1$ outside the tail even for $\xi \neq
1/6$.

To summarize this discussion the WKB condition is violated by the
present dispersion relation in the tail (region I in Fig.~\ref{fig0})
at all times and an initial vacuum state cannot be constructed
unambiguously. Outside of the tail (region II of Fig.~\ref{fig0}), for
$\omega_{\rm phys}\gg H$ or $k_{\rm phys}\ll K_+$, the WKB
approximation is a good approximation. One can also verify that the
construction of an initial vacuum state by minimization of the energy
content does not work in this case, see Ref.~\cite{LMMU01}. This point
is one major obstacle to the scenario proposed in
Ref.~\cite{MBK01}. Since there is no preferred initial vacuum state,
all cosmological conclusions drawn depend directly on the particular
choice of the initial state, hence on the choice of initial data. At
the very best, one has to fine-tune the initial conditions to obtain a
given amount of energy in a given part of the spectrum.

 The standard calculation of the amount of energy contained at late
times in a given co-moving wave-number mode is done by decomposing the
solution at late times (outside the tail) in terms of positive and
negative frequency plane waves, as

\begin{equation}\label{eq:planew}
 \mu_k^{\rm out} = \frac{\alpha_k}{\sqrt{2\omega^{\rm out}_k}}
 \hbox{e}^{-i\omega^{\rm out}_k\eta} \, + \,
 \frac{\beta_k}{\sqrt{2\omega^{\rm out}_k}}
\hbox{e}^{i\omega^{\rm out}_k\eta}\, .
\end{equation}
The squared modulus of the Bogoliubov coefficient $\beta_k$ then will give
the occupation number of quanta produced in the mode of co-moving
wave-number $k$. Note that, in principle, the coefficients $\alpha_k$
and $\beta_k$ can be slowly varying functions of time, and the above
expression implicitly involves a WKB approximation to first order in
which the time evolution of $\alpha_k$ and $\beta_k$ is neglected. The
corresponding vacuum is an adiabatic vacuum to first order.

In Ref.~\cite{MBK01} $\beta_k$ is calculated in the limit
$\eta\rightarrow +\infty$ as $\omega^{\rm out}\rightarrow
\sqrt{\epsilon_1} k$. However the limit $\eta\rightarrow+\infty$ does
not hold in an inflationary Universe with
$a\propto\vert\eta\vert^{-\beta}$ and $\beta\geq1$ since $a$ is
singular as $\eta\rightarrow0^-$. One needs to match the background
evolution to a decelerated Universe as $\eta\rightarrow 0^-$. In
effect, if one wishes to calculate the contribution of the tail modes
to the energy density {\em today}, it is necessary to calculate the
evolution of the modes from the inflationary era up to today. Note
that the dynamical evolution of the tail modes depends {\it a priori}
strongly on the background scale factor dynamics.

This calculation could not be performed in Ref.~\cite{MBK01}, since
the solution given in terms of the hypergeometric function is not
valid at late times in the radiation dominated or matter dominated
eras unless the parameter $x$ of the dispersion relation is tuned to
the evolution of the scale factor, but the dispersion relation would
become pathological as we saw before for $x=(k_{\rm phys}/k_{\rm
c})^{1/\beta}$ with $\beta<0$. Furthermore, as explained above, the
solution to the field equation given in Ref.~\cite{MBK01} [see
Eq.~(\ref{eq:4})] describes only one branch of the solution. Finally,
one cannot compute $\beta_k$ for modes still contained in the ``tail''
at late times by matching the solution to plane waves as done in
Ref.~\cite{MBK01} since for those modes, the WKB approximation is
never valid so that the out solution cannot be decomposed in a sum of
plane waves. Thus the calculation of the Bogoliubov coefficient
$\beta_k$ performed in Ref.~\cite{MBK01} cannot apply to modes
contained in the ``tail'' today.

\section{The tail energy density}\label{Sec:III}

In this Section we calculate the amount of energy density created in
quanta that redshift out of the ``tail'', and show that it leads to a
severe back-reaction problem. In Ref.~\cite{MBK01}the energy density
contained in the tail is calculated as
\begin{equation}\label{eq:7}
 \langle\rho_{\rm tail}\rangle = \frac{1}{2\pi^2}
\int_{k_{\rm H}}^{+\infty}
 k_{\rm phys}{\rm d}k_{\rm phys}\int \omega_{\rm phys} {\rm
 d}\omega_{\rm phys} \vert\beta_{k_{\rm phys}}\vert^2,
\end{equation}
where $k_{\rm H}$ is the physical wave-number such that $\omega_{\rm
H}\equiv\omega_{\rm phys}(k_{\rm phys})=H_0$ today. This expression
for $\langle\rho_{\rm tail}\rangle$ is ill-defined due to the double
integration element ${\rm d}k\,{\rm d}\omega$ in the absence of a
Dirac function on the mass shell. The total energy density
$\langle\rho_{\rm total}\rangle$ is defined analogously but the lower
bound is extended to $k=0$. Then, it is argued that $\langle\rho_{\rm
tail}\rangle/\langle\rho_{\rm total}\rangle \simeq 10^{-122}$ during
inflation. Note that if $\rho_{\rm tail}\simeq 10^{-122}M_{{\rm
Pl}}^4$ and $\langle\rho_{\rm tail}\rangle$ is constant (corresponding
to a acuum-like equation of state as suggested) in order to account
for the dark energy then the above statement yields $\rho_{\rm
total}\sim M_{{\rm Pl}}^4$. If this holds during inflation, one faces
a severe back-reaction problem since the background energy density
during inflation is $\sim10$ orders of magnitude below $M_{{\rm
Pl}}^4$, and the overall calculation framework (a test quantum scalar
field on a classical background) breaks down. As we argue in this
Section, it is actually a generic prediction of this model that
$\rho_{\rm total}\sim M_{{\rm Pl}}^4$ {\it at all times}. This result
$\rho_{\rm total}\sim M_{{\rm Pl}}^4$ is in agreement with a recent
work by Starobinsky~\cite{S01}, which showed that models with
dispersion relations such that the WKB approximation is not valid in
the far past when the physical wave-number $k_{\rm phys}\gg M_{{\rm
Pl}}^4$ implies a very substantial amount of particle production.

In the following we calculate the amount of energy density stored in
modes with physical wave-number $k_{\rm phys}\sim M_{{\rm Pl}}$. The
calculation follows the line of thought indicated in the previous
Section. Since in the range $H\ll k_{\rm phys}\ll K_+$ the WKB
approximation is valid, one can decompose the solution to the field
equation in terms of plane waves as in Eq.~(\ref{eq:planew}) when
modes enter this regime. As long as $\vert Q/\Omega_k^2\vert\ll1$ one
can neglect the time evolution of $\beta_k$, and it is natural to
interpret $\vert\beta_k\vert^2$ as the occupation number of particles
in mode $k$.  As argued earlier this decomposition in plane waves
cannot be made for modes that are still contained in the tail.

One can then calculate the amount of energy density ${\rm
d}\rho_\omega/{\rm d}\ln(k_{\rm phys})$ stored in the log interval
around the physical wave-number $k_{\rm phys}$ and the corresponding
fractional density parameter ${\rm d}\Omega_{\omega}/{\rm
d}\ln(k_{\rm phys})$ in units of the background energy density:

\begin{equation}\label{eq:domega}
\frac{{\rm d}\Omega_\omega}{{\rm d}\ln k_{\rm phys}} =
\frac{4}{3\pi}\frac{k_{\rm phys}^3 \omega_{\rm phys}}{H^2 M_{\rm
Pl}^2}\vert\beta_k\vert^2,
\end{equation}
using ${\rm d}\rho_\omega/{\rm d}\ln(k_{\rm phys})=k_{\rm
phys}^3\omega_{\rm phys}\vert\beta_k\vert^2/2\pi^2$. The fractional
density parameter must be smaller than unity {\it at all times} and
{\it for all physical wave-numbers}, otherwise back-reaction is
significant and all semi-classical first order calculations are
unreliable.  In the following we calculate this quantity ${\rm
d}\Omega_{\omega}/{\rm d}\ln(k_{\rm phys})$ for a physical wave-number
$k_{\rm phys}\sim k_{\rm c}$, i.e., once the wavelength becomes larger
than the fundamental scale. It can be expressed {\it via} $\beta_k$ in
terms of the constants that parametrize the choice of initial
data. Our goal here is to study the dependence of the amount of energy
density created in modes of physical momenta $\sim M_{{\rm Pl}}$ on
the initial data, for which there is no definite prescription as we
argued in the previous Section.

\subsection{Conformal coupling: $\xi=1/6$}

In the case of conformal coupling $\xi=1/6$ there exists an exact
solution to the field equation written in terms of the two independent
solutions $\mu^{(1)}_k$ and $\mu^{(2)}_k$ in Eq.~(\ref{eq:mu1mu2}).
One can then calculate the Bogoliubov coefficient deep in the region
where the WKB approximation is valid, for instance around $k_{\rm
phys}\sim k_{\rm c}$ by decomposing this exact solution in plane
waves. However the coefficients of the hypergeometric function in term
of which the exact solution hence $\beta_k$ are written are of order
$k_{\rm c}/H\gg 1$. For values of these coefficients that are relevant
for our cosmological applications (i.e. $k_{\rm c}/H\sim 10^6$ during
inflation), the numerical calculation of the hypergeometric function
turns out to be too involved and we have been unable to calculate
$\beta_k$ in a reasonable amount of time for $k_{\rm c}/H\gtrsim
10^3$. Therefore we take a different approach and approximate the
exact solution in the tail $k_{\rm phys}>K_+$ by the solution derived
in terms of Bessel functions in Eq.~(\ref{footeq}), and that in the
region $k_{\rm c}< k_{\rm phys} < K_+$ by the plane wave solution. The
Bogoliubov coefficient $\beta_k$ of the plane wave solution is
obtained by matching the two solutions and their first derivatives at
the transition point $k_{\rm phys}=K_+$. Of course, it gives an
approximation to $\beta_k$, but as we show in the following the
deviation from the overall behavior of $\beta_k$ away from its minimum
and on its behavior around its minimum are negligible. We thus proceed
as follows: in the following Section, we calculate the Bogoliubov
coefficient denoted $\beta_k^{\rm (approx)}$ by solving for the Bessel
functions in the remote past and performing the matching at $K_+$. In
the subsequent Section, we calculate the Bogoliubov coefficient
$\beta_k^{\rm (exact)}$ analytically using the exact solution and
demonstrate that $\beta_k^{(\rm approx)}$ is a good approximation for
values of $k_{\rm c}/H$ as high as $\simeq 10^3$. Finally we examine
the behavior of $\beta_k^{\rm (approx)}$ and evaluate the amount of
energy density produced by the non-adiabatic evolution of modes in the
``tail'' for realistic values of $k_{\rm c}/H$. This calculation is
entirely analytical, only the verification of the accuracy of the
approximation is numerical.

\subsubsection{Approximate calculation of the Bogoliubov coefficient}

As already mentioned above, see Eq.~(\ref{footeq}), the mode function
in the tail can be approximatively expressed in terms of the Bessel
and Neumann functions $J_0$ and $N_0$ as
\begin{equation}
\label{solJN}
\mu _k(\eta )\simeq \frac{1}{\sqrt{2k}}\biggl[A_1(k)J_0({\rm e}^y)
+A_2(k)N_0({\rm e}^y)\biggr]\, ,\quad y\equiv \frac{H}{2k_{{\rm c}}}k\eta 
+\frac12\ln \biggl[\frac{4k_{{\rm c}}^2(4-\epsilon _1)}{H^2}\biggr]\, .
\end{equation}
This solution is valid if the scale factor is that of the de Sitter
space-time: $a(\eta )=-1/(H\eta )$. The mode function must satisfy the
relation $W\equiv \mu _k\mu _k^*{}'-\mu _k\mu _k^*=i$. Using the above
equation, one finds that the Wronskian is equal to
$W=-H(A_2A_1^*-A_1A_2^*)/(2\pi k_{{\rm c}})$. As a consequence, if one
represents the coefficient $A_2$ in polar form, $A_2\equiv r{\rm
e}^{i\Phi }$, one has $A_1=-\pi k_{{\rm c}}/(Hr\sin \Phi )$, where we
have chosen $A_1$ to be real. The parameters $r$ and $\Phi$ will
characterize in the following the choice of initial data.

In the region where the WKB approximation is valid, i.e. for $\omega
_{{\rm phys }}\gg H$, one has
\begin{equation} 
\mu _k(\eta )\simeq \frac{\alpha _k}{\sqrt{2\omega (k,\eta )}}
{\rm e}^{-i\Omega }+\frac{\beta _k}{\sqrt{2\omega (k,\eta )}}
{\rm e}^{i\Omega }\, ,
\end{equation}
where $\Omega \equiv \int ^{\eta }{\rm d}\tau \omega (k,\tau )$. In
order to express the Bogoliubov coefficient $\vert \beta _k\vert $ in
terms of the constants parameterizing the choice of the initial data in
the tail, $A_1(k)$ and $A_2(k)$, we use the continuity of the mode
function $\mu _k$ and of its derivative at the transition between the
two regions at $y=y_{\rm m}$, for which $\omega _{{\rm phys }}(y_{\rm
m})=\sqrt{2}H$. The result reads
\begin{equation}\label{eq:beta-approx1}
\beta _k^{(\rm approx)} = \frac{i{\rm e}^{-i\Omega }}{\sqrt{4k\omega
(k,y_{\rm m})}} \biggl\{A_1(k)\biggr[-\gamma _k(y_{\rm m})J_0({\rm
e}^{y_{\rm m}})+\frac{H}{2}\frac{k}{k_{{\rm c}}} {\rm e}^{y_{\rm
m}}J_1({\rm e}^{y_{\rm m}})\biggr]+ A_2(k)\biggr[-\gamma _kN_0({\rm
e}^{y_{\rm m}}) +\frac{H}{2}\frac{k}{k_{{\rm c}}} {\rm e}^{y_{\rm
m}}N_1({\rm e}^{y_{\rm m}}) \biggr]\biggr\} \, ,
\end{equation}
where $\gamma _k\equiv \omega '/(2\omega )+i\omega $. Working out this 
last expression, one obtains
\begin{equation}
\label{betaka12}
\beta_k^{(\rm approx)} = \frac{i{\rm e}^{-i\Omega }}{\sqrt{4\sqrt{2}}}
\biggl(\frac{H}{K_+}\biggr)^{1/2}
\biggl\{A_1(k)\biggr[-\biggr(\frac{K_+}{4k_{{\rm c}}}+i\sqrt{2}\biggl)
J_0({\rm e}^{y_{\rm m}})+\sqrt{2}J_1({\rm e}^{y_{\rm m}})\biggr]+
A_2(k) \biggr[-\biggr(\frac{K_+}{4k_{{\rm c}}}+i\sqrt{2}\biggl)
N_0({\rm e}^{y_{\rm m}})+\sqrt{2}N_1({\rm e}^{y_{\rm
m}})\biggr]\biggr\}\, .
\end{equation}
We are now in a position where we can compute the $\vert \beta _k^{(\rm approx)}\vert
^2$ using the parametrization of the coefficients $A_1$ and $A_2$
introduced above. The final result reads
\begin{equation}
\label{betakrphi}
\vert \beta _k^{(\rm approx)}\vert ^2 (\rho ,\Phi )=\frac{\pi
^2}{4\sqrt{2}}\frac{J}{\rho ^2\sin ^2\Phi } +\frac{N\rho
^2}{4\sqrt{2}} -\frac{\pi K}{2\sqrt{2}}{\rm cot}\Phi -\frac12 \,,
\end{equation}
where we have defined the rescaled variable $\rho $ by $\rho \equiv
r\sqrt{H/K_+}$ and where the coefficients $J$, $N$ and $K$ can be
expressed as
\begin{eqnarray}
J &=& \frac{1}{16}J_0^2-\frac{1}{\sqrt{2}}\frac{k_{{\rm
c}}}{K_+}J_0J_1+2\frac{k_{{\rm c}}^2}{K_+^2}(J_0^2+J_1^2)\, ,
\quad 
N = \frac{K_+^2}{16k_{{\rm c}}^2}N_0^2
+\frac{1}{\sqrt{2}}\frac{K_+}{k_{{\rm c}}}N_0N_1+2(N_0^2+N_1^2)\, ,
\\
K &=& \frac{K_+}{16k_{{\rm c}}}J_0N_0
+\frac{\sqrt{2}}{4}(J_0N_1+J_1N_0)+2\frac{k_{{\rm c}}}{K_+}(J_0N_0
+J_1N_1)\, .
\end{eqnarray}
The Bessel and Neumann functions are evaluated at the matching point
for which their argument reads ${\rm e}^{y_{\rm m}}=2\sqrt{2}k_{{\rm
C}}/K_+$. A direct calculation shows that $JN-K^2=2/\pi ^2$. The
Bogoliubov coefficient $\vert \beta _k^{(\rm approx)}\vert ^2$ can be viewed as a
two-dimensional surface parametrized by the polar coordinates $(\rho
,\Phi )$.

\subsubsection{Test of the method of approximation}

Before studying the above Bogoliubov coefficient in greater detail,
one must check that the approximation is well-controlled. For this
purpose, it is interesting to calculate the Bogoliubov coefficient
using the exact solution expressed in terms of hypergeometric
functions
\begin{equation}
\label{exact}
\mu_k(\eta) = \frac{1}{\sqrt{2k}}\left[C_1(k) \mu_k^{(1)}(\eta) +
C_2(k) \mu_k^{(2)}\right],
\end{equation}
where the functions $\mu_k^{(1)}$ and $\mu_k^{(2)}$ have been defined
in Eq.~(\ref{eq:mu1mu2}) above, and the (dimensionless) functions
$C_1(k)$ and $C_2(k)$ are related to each other by the Wronskian
normalization condition~\cite{W02}:
\begin{equation}\label{eq:wronskian-exact}
\mu_k\mu_k^*{}' - \mu_k^\prime\mu_k^*= \left[\vert C_1(k)\vert^2 -
\vert C_2(k)\vert^2\right] \frac{{\rm e}^{2i\pi b}}{2k_{\rm
c}\vert\eta_0\vert}{\cal F},
\end{equation}
with the numerical factor ${\cal F}$ is written in terms of the
parameters $b$ and $d$, as:
\begin{eqnarray}
{\cal F} & = & \frac{\left(b+d +
\frac{1}{2}\right)^2}{2b+1}{}_2F_1\left(b+d+\frac{3}{2}, b-d+{1\over
2};2b+2;-1\right){}_2F_1\left(-b-d+{1\over2},-b+d+{1\over2};-2b+1;-1\right)  + 
\nonumber\\
 & &  2b\,{}_2F_1\left(b+d+{1\over2},b-d+{1\over2};2b+1;-1\right)
{}_2F_1\left(-b-d+{1\over2},-b+d+{1\over2};-2b+1;-1\right)  + 
\nonumber\\
 & & \frac{\left(-b+d+{1\over2}\right)^2}{2b-1}
{}_2F_1\left(b+d+{1\over2},b-d+{1\over2};2b+1;-1\right)
{}_2F_1\left(-b-d+{1\over2},-b+d+{3\over2};-2b+2;-1\right).
\end{eqnarray}
This solution is valid at all times since it is an exact solution of
the field equation. In this case, one can calculate the Bogoliubov
coefficient at any time provided the WKB approximation is then valid, using:
\begin{equation}
\label{betaexact}
\vert \beta _k ^{\rm (exact)}\vert =\frac{1}{\sqrt{2\omega }} \biggl \vert \mu
_k'+\biggl(\frac{\omega '}{2\omega }+i\omega \biggr)\mu _k \biggr
\vert\, ,
\end{equation}
where, in the last expression, $\mu _k$ is given by
Eq.~(\ref{exact}). Notice that this procedure differs from the
previous calculation of the Bogoliubov coefficient. Here, we do not
perform a matching at the transition between the tail and the WKB
region but rather use the exact solution (\ref{exact}) all the way
through and calculate its ``overlap'' with the WKB solution deep in
the WKB region. The initial conditions enter this expression via the
two constants $C_1(k)$ and $C_2(k)$.

\par

\begin{figure}
      \centering
      \includegraphics[width=0.49\textwidth,clip=true]{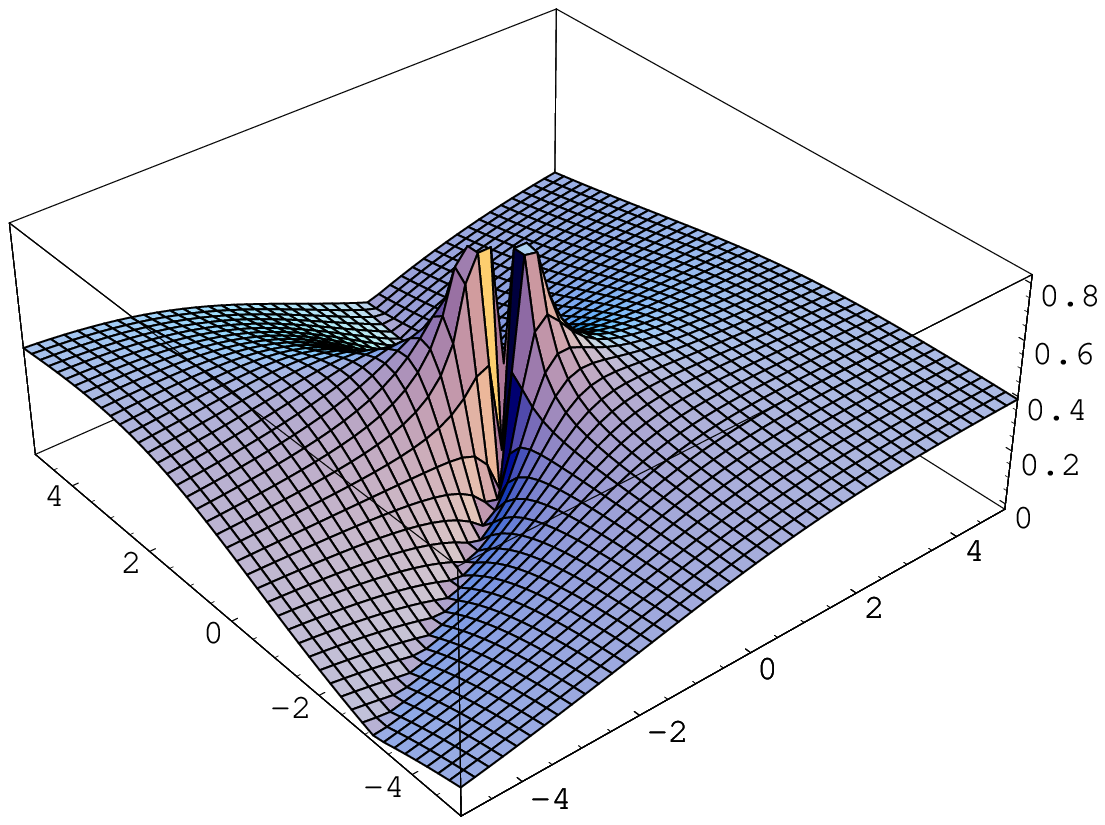}     
      \includegraphics[width=0.49\textwidth,clip=true]{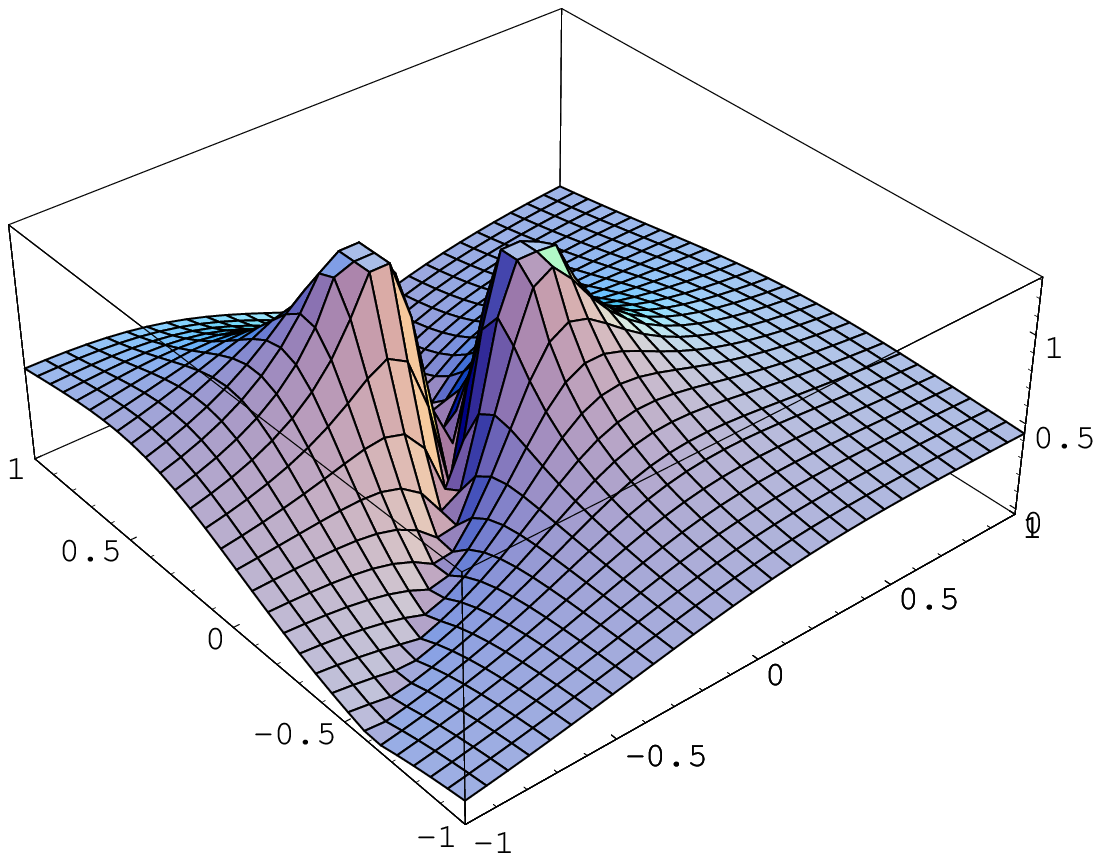}
      \caption[...]{Left panel: The function $A$ defined in
      Eq.~(\ref{error}) that quantifies the deviation of the approximate
      $|\beta _k|$ with respect to its exact form. $A$ is depicted as
      a function of $(R,\theta)$ that parameterize the initial
      conditions ($C_1,C_2$) of the exact solution.  Right panel: zoom
      around the origin in order to unveil the two peaks structure.}
\label{3d}
\end{figure}

We need to compare $\vert \beta _k ^{\rm (exact)}\vert $ with $\vert
\beta _k ^{\rm (approx)}\vert $ for the same initial conditions. Since
$\vert \beta _k ^{\rm (approx)}\vert $ is expressed in terms of the
constants $A_1(k)$ and $A_2(k)$, one needs to re-express $A_1(k)$ and
$A_2(k)$ in terms of $C_1(k)$ and $C_2(k)$. This can be done by
matching the asymptotic behaviors of the two solutions deep in the
tail, i.e. in the limit $\eta\rightarrow -\infty$. There, the
approximate solution given by Eq.~(\ref{solJN}) reduces to
\begin{equation}
\mu _k(\eta )\simeq \frac{1}{\sqrt{2k}}\biggl\{A_1 -A_2\frac{H}{\pi
k_{{\rm c}}}k\vert \eta \vert -\frac{2A_2}{\pi }\ln 2+\frac{2A_2}{\pi
}\gamma _{{\rm E}} +\frac{A_2}{\pi }\ln \biggl[\frac{4k_{{\rm
c}}^2(4-\epsilon _1)}{H^2}\biggr]\biggr\}\, ,
\end{equation}
where $\gamma _{{\rm E}}$ is the Euler constant, $\gamma _{{\rm
E}}\simeq 0.5772$. On the other hand, the exact solution of
Eq.~(\ref{exact}) can be written as
\begin{eqnarray}
\mu _k(\eta ) &\simeq &\frac{1}{\sqrt{2k}}\frac{H}{k_{{\rm c}}}k\vert
\eta \vert\biggl\{(G C_1 + G^*C_2) -
G C_1\biggl[2\gamma _{{\rm E}} +\Psi \biggl(b+d+\frac12\biggr)+\Psi
\biggl(b-d+\frac12\biggr )\biggr] 
\nonumber \\ 
& & -G^*C_2\biggl[2\gamma _{{\rm E}} +\Psi \biggl(-b+d+\frac12\biggr)
+\Psi \biggl(-b-d+\frac12\biggr )\biggr] \biggr\}\, ,
\end{eqnarray}
where the coefficient $G$ is given in terms of the Euler Beta function
as: $G=1/B(b+d+1/2,b-d+1/2)=\Gamma (2b+1)/[\Gamma (b+d+1/2)\Gamma
(b-d+1/2)]$, and satisfies, since $b$ is pure imaginary and $d$ is
real, $G^*=G(b\leftrightarrow -b)$. This relation stems from the
asymptotic behavior of the hypergeometric functions for large values
of their argument given by Eq.~(15.3.13) of Ref.~\cite{AS}. The
di-Gamma function $\Psi (x)$ function is defined by $\Psi (x)\equiv
{\rm d}\ln \Gamma (x)/{\rm d}x$. If one identifies the constant term
and the linear term in conformal time of the two previous relations,
we obtain:
\begin{eqnarray}
A_1(k) &=& G C_1(k)\biggl\{ \ln \biggl[\frac{k_{{\rm c}}^2(4-\epsilon
_1)}{H^2}\biggr] -\Psi \biggl(b+d+\frac12 \biggr)-\Psi
\biggl(b-d+\frac12 \biggr ) \biggr\} 
\nonumber \\ 
& & + G^*C_2(k) \biggl\{ \ln \biggl[
\frac{k_{{\rm c}}^2(4-\epsilon _1)}{H^2}\biggr]
-\Psi \biggl(-b+d+\frac12 \biggr)-\Psi \biggl(-b-d+\frac12 \biggr )
\biggr\}\, , 
\\ 
A_2(k) &=& -\pi [G C_1(k) + G^*C_2(k)]\, .
\end{eqnarray} 
Then, it is sufficient to use the above relations in
Eq.~(\ref{betaka12}) to obtain $\vert \beta _k ^{\rm (approx)}\vert $
in terms of $C_1(k)$ and $C_2(k)$ and compare it to
$\vert\beta_k^{(\rm exact)}\vert$. In order to characterize the
accuracy with which the Bogoliubov coefficient is calculated, we plot
the following quantity
\begin{equation}
\label{error}
A\equiv 2 \biggl\vert \frac{\vert \beta _k ^{\rm
(approx)}\vert -\vert \beta _k ^{\rm (exact)}\vert }{\vert \beta _k
^{\rm (approx)}\vert +\vert \beta _k ^{\rm
(exact)}\vert }\biggr \vert
\end{equation}
for various values of the coefficients $C_1(k)$ and $C_2(k)$. More
precisely, we use a polar representation and take $C_2(k)=R{\rm
e}^{i\theta }$ while $C_1(k)$ is real and calculated in terms of
$C_2(k)$ using the Wronskian relation Eq.~(\ref{eq:wronskian-exact}). In
Fig.~\ref{3d}, we have plotted $A(R,\theta )$ for $k_{{\rm c}}/H=10^2$
and $k_{{\rm c}}=M_{{\rm Pl}}$.
We see that the error for large values of $\rho$ is less than
$\sim40\,$\% and is constant, i.e. the offset between the two
Bogoliubov coefficients does not depend on $\rho$ and $\Phi$ in a
first approximation. For $\rho\sim0$, the error increases to 1; this
artefact comes from the fact that the minima where the two Bogoliubov
coefficients vanish are slightly offset one from the other. If one
coefficient vanishes while the other remains finite and non-zero, then
the value of $A$ is pushed toward one, $A\rightarrow 1$. This error
however is of no consequence for what follows. Indeed we will not be
interested in the location of the minimum but in the behavior of
$\beta_k$ around the minimum and far from the minimum. As is obvious
from Fig.~\ref{3d}, these behaviors match closely in both cases and
our approximation will be sufficient for our purposes.  We have
checked that the function $A$ remains the same for other values of
$k_{{\rm c}}/H$ which allow numerical calculations, i.e. $k_{{\rm
c}}/H\in [10, 10^{3}]$.

\par

The situation is in fact very similar to the standard calculation of
the power spectrum in an inflationary theory: in principle, one cannot
match two different branches at a point where the approximation breaks
down (for standard inflation this occurs at first horizon
crossing). However, since the approximation is only violated in a
small region one expects the corresponding result to be correct at
leading order. This is indeed the case for inflation for which the
amplitude of the spectrum is predicted up to a factor of order unity
and the spectral slope is unchanged by the matching. Here we also find
that $ \vert \beta _k ^{\rm (approx )}\vert ={\cal O}(1)\vert \beta _k
^{\rm (exact)}\vert $.

\subsubsection{Fine-tuning of the initial conditions}

Since we have demonstrated that the approximation to
$\vert\beta_k\vert$ is quite reasonable, we now study the behavior of
Eq.~(\ref{betakrphi}) for more realistic values of the ratio $k_{{\rm
c}}/H$. The Bogoliubov coefficient possesses an absolute minimum with
$\beta_k=0$ located at
\begin{equation}
\rho _{\rm min}^4=\frac{\pi ^2J^2}{JN-K^2}\, , \quad 
\cos \Phi _{\rm min}=\frac{K}{\sqrt{JN}}\, ,
\end{equation}
using the notations defined previously. One should not be surprised to
find a minimum with $\beta_k=0$ since one can express the matching of
the two branches of the solution and their first derivatives at
$\eta_{\rm m}$ as two equations relating the coefficients $A_1(k)$ and
$A_2(k)$ as a function of $\alpha_k$ and $\beta_k$ and find a solution
with $\beta_k=0$. The Wronskian normalization condition is always
satisfied by both branches of the solution. This solution with initial
conditions ($\rho_{\rm min}$, $\Phi_{\rm min}$) corresponds to a choice
of initial data such that at late times, when modes have exited from
the tail, their quantum state is that of an adiabatic vacuum. Note
therefore that there is no naturalness in choosing these initial
conditions since the adiabatic vacuum is only a late time consequence
of such initial data. Furthermore one can show that for generic
initial data, the state of the quantum field at late times is not an
adiabatic vacuum, hence quanta have been produced.

 Indeed the behavior of $\vert \beta_k \vert ^2$ around this absolute
minimum can easily be established. From a Taylor expansion, one
obtains
\begin{equation}
\vert \beta _k \vert ^2 \simeq \frac{N}{\sqrt{2}}(\rho -\rho _{\rm
min})^2\, , \quad \vert \beta _k \vert ^2 \simeq \frac{\pi
^4}{16}J^2N^2(\Phi -\Phi _{\rm min})^2\, .
\end{equation}
For a crude order of magnitude estimate, one can develop the Bessel
functions to first order in the small argument limit $e^{y_{\rm
m}}=2\sqrt{2}k_{{\rm c}}/K_+\ll 1$ (more exactly, for de Sitter
inflation and $k_{\rm c}=M_{\rm Pl}$, one has $k_{\rm c}/K_+\simeq
0.06$). This leads to $J\simeq1/16 +{\cal O}(k_{{\rm c}}^2/K_+^2)$ and
$N\simeq \ln ^2(\sqrt{2}k_{\rm c}/K_+)/(4\pi ^2)(K_+^2/k_{{\rm
c}}^2)+{\cal O}(k_{{\rm c}}^0/K_+^0)$. Thus in order to avoid a
back-reaction problem, the initial conditions $\rho$ and $\Phi$ must
not differ too much from $\rho_{\rm min}$ and $\Phi_{\rm min}$ which
lead to $\beta_k=0$ (hence a zero amount of energy density created).
More precisely, the energy density produced is of the order of the
background energy density, i.e. ${\rm d}\Omega_\omega/{\rm
d}\ln(k_{\rm phys})=1$, when $\rho$ or $\Phi$ respectively depart from
the minimum by amounts $\delta\rho$ or $\delta\Phi$ given by:
\begin{equation}
\delta \rho \simeq {\cal O}\biggl[\frac{H}{M_{\rm Pl}}\ln ^{-1}\biggl(
\frac{M_{\rm Pl}}{H}\biggl)\biggr]\, , \quad \delta \Phi \simeq {\cal
O}\biggl[\frac{H}{M_{\rm Pl}}\ln ^{-2}\biggl( \frac{M_{\rm
Pl}}{H}\biggl)\biggr]\, .
\end{equation}
Here we assumed $k_{\rm c}=M_{\rm Pl}$. Hence the corresponding
fine-tuning of the initial conditions is of order $H/M_{\rm Pl}$ (if
one assumes a uniform measure in $\rho$ and $\Phi$ in parameter
space).

One should note that the above constraint is valid for a given
co-moving wave-number $k$ and has been calculated at a time when
$k/a=k_{\rm c}$. Since the fractional density parameter of quanta
extracted out of the vacuum ${\rm d}\Omega_\omega/{\rm d}\ln(k_{\rm
phys})$ must be smaller at all times during inflation, i.e. for a
range of co-moving wave-numbers $k$ since $k$ and $\eta$ can be related
by the above constraint $k/a=k_{\rm c}$, the above constraint on
$\rho_{\rm min}$ and $\Phi_{\rm min}$ rather applies to a continuum of
values of co-moving wave-numbers. In other words one does not have to
fine-tune two parameters characterizing the initial data but a whole
continuum of parameters, i.e. the functions $\rho_{\rm min}(k)$ and
$\Phi_{\rm min}(k)$ themselves. The dependence in $k$ of these
functions is hidden in the argument of the Bessel and Neumann functions
${\rm e}^{y_{\rm m}}$, since $y_{\rm m}$ depends on $k$.

\subsection{Non conformal coupling: $\xi\not=1/6$}

One can also perform a similar calculation of the Bogoliubov
coefficient when the coupling is no longer conformal $\xi\neq1/6$. In
this case the calculation can be performed analytically for all
background scale factors. For the sake of simplicity we choose minimal
coupling $\xi=0$ but this can be trivially expanded to various choices
of the coupling to gravity, and does not modify the conclusions we
derive below.

If $\xi=0$, the evolution of the modes is dominated by $a''/a$ in the
tail, i.e.  when $\omega_{\rm phys}\ll H$ ($k_{\rm phys}\gg k_{\rm
c}$), and the solution can be written as:

\begin{equation}\label{eq:mu-a}
\mu_k(\eta)\,=\,\frac{1}{\sqrt{2k}}\left[C_+(k) \frac{a(\eta)}{a_{\rm
i}} - C_-(k)\frac{a_{\rm i}}{\eta_{\rm i}} a(\eta)\int_{\eta_{\rm
i}}^{\eta}\frac{{\rm d}\tau}{a^2(\tau)}\right],
\end{equation}
where $C_+(k)$ and $C_-(k)$ are two dimensionless $k-$dependent
constants, and $\eta_{\rm i}$ is some initial conformal time. One can
check that this solution and the power-laws given in
Eq.~(\ref{eq:power-law}) are the same. Here one cannot choose the time
of matching $\eta_{\rm m}$ to the WKB solution, since the matching has
to be done when $\Omega_k=0$, i.e. when $\omega_k^2 = a''/a$ or
$k_{\rm phys}=K_+$. In the region $\omega_{\rm phys}\gg H$, i.e. for
$\eta \gg \eta_{\rm m}$, the WKB approximation is valid as we have
seen before, and the matching to the WKB form is justified at
$\eta=\eta_{\rm m}(k)$. For a given wave-number $k$, we are free to
set $\eta_{\rm i}=\eta_{\rm m}$, since this amounts to a redefinition
of the constants $C_+(k)$ and $C_-(k)$ by a function of $k$. The
matching at $\eta_{\rm m}$ then gives:

\begin{equation}\label{eq:alpha-beta-mu-a}
\alpha_k \, = \, \frac{i}{\sqrt{4k\omega_k}}\left[
C_+ \biggl(\gamma_k^*+\frac{a'}{a}\biggr)- 
\frac{C_-}{\eta_{\rm m}}\right],
\quad\quad
\beta_k \,  =  \, \frac{-i}{\sqrt{4k\omega_k}}\left[
C_+ \biggl(\gamma_k +\frac{a'}{a}\biggr)
-\frac{C_-}{\eta_{\rm m}}\right]\, ,
\end{equation}
with the function $\gamma_k\equiv (\omega'/2\omega) + i\omega$ as
above [see Eq.~(\ref{eq:beta-approx1})] and where all quantities in
the above two equations are understood to be taken at $\eta=\eta_{\rm
m}(k)$.  In particular, at time $\eta_{\rm m}$,
$\omega_k^2=a''/a=(1+\beta)a_{\rm m}^2H_{\rm m}^2/\beta$. Since
$k_{\rm phys}\gg k_{\rm c}$ at $\eta_{\rm m}$ one can use the limiting
form of $\omega_{\rm phys}$ given in Eq.~(\ref{eq:omega-approx}),
hence:

\begin{equation}
\gamma_k +\frac{a'}{a} \simeq a_{\rm m}H_{\rm
m}\left(\frac{K_+}{4k_{\rm c}} + 1 +
i\sqrt{\frac{1+\beta}{\beta}}\right).
\end{equation}

The constants $C_+(k)$ and $C_-(k)$ are related to one another by the
normalization of the mode functions:
$\mu_k\mu_k^{*\prime}-\mu_k^\prime\mu_k^*=i$, which gives:
\begin{equation}
C_+(k)\,=\,-\beta \frac{K_+}{H_{\rm m}r\sin \Phi},
\end{equation}
and as before we keep $r\equiv\vert C_-(k)\vert$ and $\Phi={\rm
arg}[C_-(k)]$ as the two independent parameters characterizing the
choice of initial data.  One finally derives the squared modulus of
the Bogoliubov coefficient $\beta_k$ as:
\begin{equation}\label{eq:beta-mu-a}
\vert\beta_k\vert^2 \,= \,
\frac{1}{r^2\sin^2\Phi}\frac{\beta^2}{4}\sqrt{\frac{\beta}{\beta+1}}
\frac{K_+}{H_{\rm
m}}\left[\left(1 + \frac{K_+}{4k_{\rm c}}\right)^2 +
\frac{1+\beta}{\beta}\right]
+\frac{r^2}{4\beta ^2}\sqrt{\frac{\beta}{1+\beta}}\frac{H_{\rm m}}{K_+}
-\frac12\sqrt{\frac{\beta}{1+\beta}}\left(1+\frac{K_+}{4k_{\rm
c}}\right){\rm cot}\Phi - \frac{1}{2}.
\end{equation}
Since $K_+/H_{\rm m}$ is a large number, in the following we use the
rescaled variable $\rho\equiv r\sqrt{H/K_+}$ instead of
$r$. Equation~(\ref{eq:beta-mu-a}) above is particularly attractive
because it has exactly the same functional shape as
Eq.~(\ref{betakrphi}). It allows us to understand analytically the
behavior of the amount of energy density produced in modes with
$k_{\rm phys}\sim k_{\rm c}$ as a function of the initial data. The
occupation number $\vert\beta_k\vert^2 $ has an absolute minimum
located at:
\begin{equation}
\rho _{\rm min}=\vert \beta \vert \biggl(\frac{\beta }{1+\beta }\biggr)^{1/4}
\sqrt{\biggl(1+\frac{K_+}{4k_{\rm c}}\biggr)^2+\frac{1+\beta }{\beta }}
\, , \quad 
\cos\Phi_{\rm min}\,=\,\frac{1+K_+/(4k_{\rm c})}
{\sqrt{\left[1+K_+/(4k_{\rm c})\right]^2 +
(1+\beta )/\beta}}.
\end{equation}
As before the occupation number vanishes exactly at this minimum, but
the back-reaction problem cannot be avoided for generic initial
conditions. In the present case it is not possible to make a sensible
contour plot of ${\rm d}\Omega_\omega/{\rm d}\ln(k_{\rm phys})$ since
this function changes by many orders of magnitude over very small
intervals of $\rho, \Phi$. Therefore, we take a conservative approach
in which we calculate $\vert\beta_k(\rho )\vert^2$ as a function of
$\rho $ for the values of $\Phi $ that minimize this quantity at each
$\rho $. We also evaluate $\vert\beta_k(\Phi)\vert^2$ as a function of
$\Phi$ for the values of $\rho$ that minimize this quantity at each
$\Phi$. In other words, we solve $\partial_{\Phi
}\vert\beta_k\vert^2=0$ for $\Phi $ as a function of $\rho $ and
$\partial_\rho \vert\beta_k\vert^2=0$ for $\rho$ as a function of
$\Phi$:
\begin{equation}
\Phi _{\rm min}(\rho )=\tan ^{-1}\biggl\{\frac{\beta ^2}{\rho ^2}
\frac{[1+K_+/(4k_{\rm c})]^2+(1+\beta)/\beta }{1+K_+/(4k_{\rm c})}\biggr\}
\, ,\quad  
\rho_{\rm
min}^4(\Phi)\,=\,\frac{\beta ^4}{\sin^2\Phi}\left[\left(1+\frac{K_+}{4k_{\rm
c}}\right)^2 + \frac{1+\beta}{\beta}\right],
\end{equation}
\begin{figure}
      \centering
      \includegraphics[width=0.49\textwidth,clip=true]{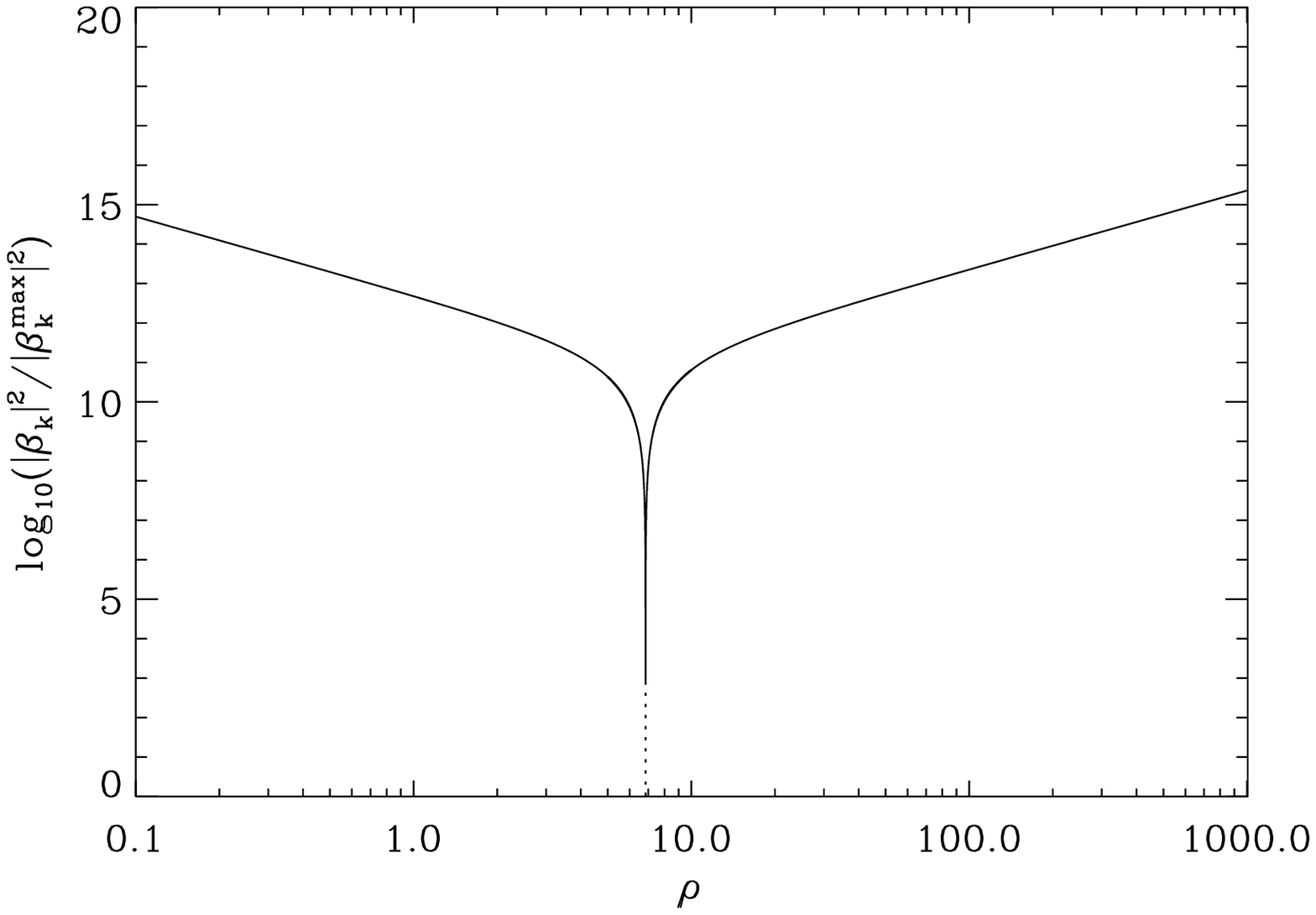}
      \includegraphics[width=0.49\textwidth,clip=true]{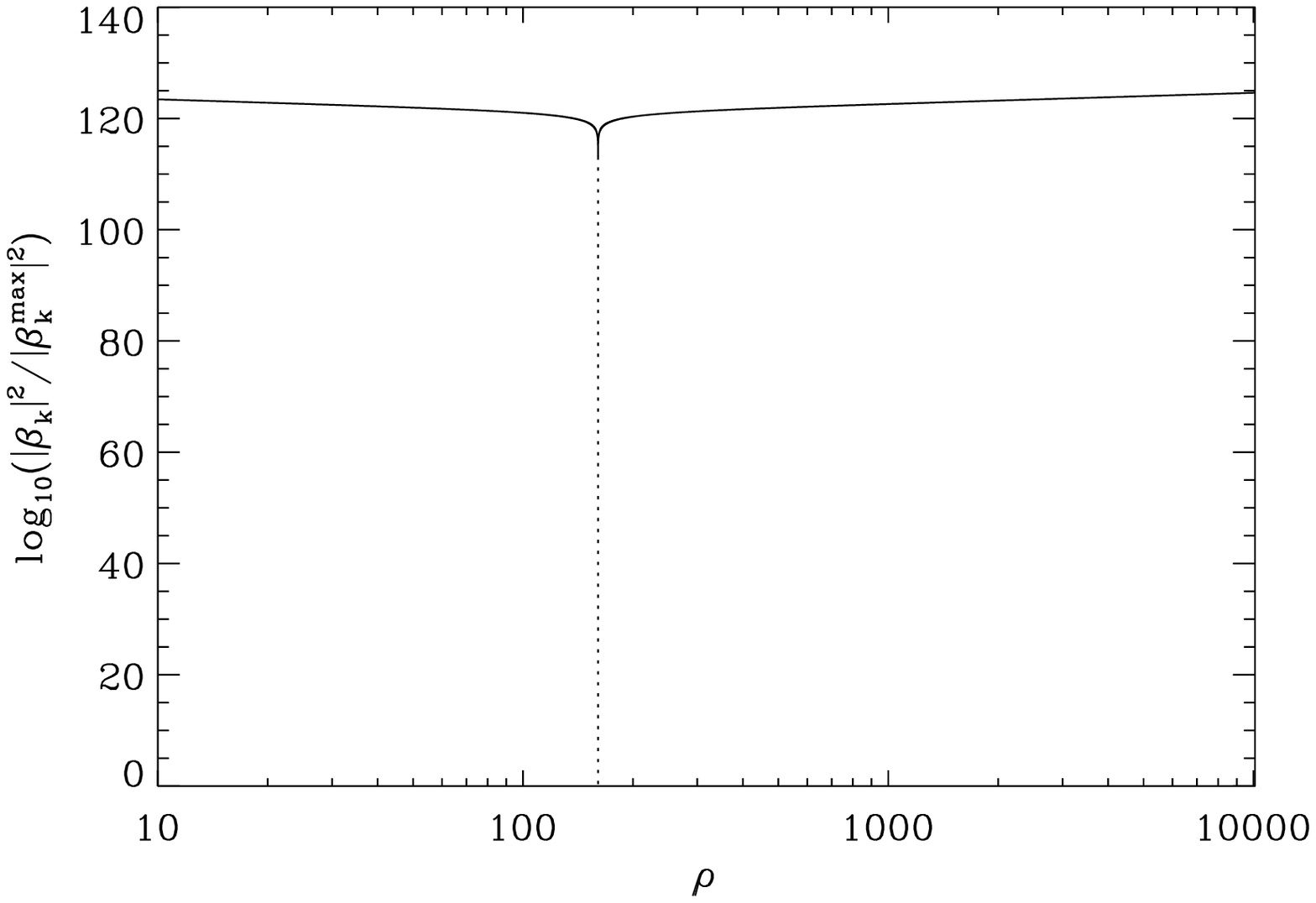}
      \caption[...]{Left panel: the solid line represents
      $\log_{10}\left(\vert\beta_k\vert^2 / \vert\beta_k^{\rm
      max}\vert^2\right)\equiv \log_{10}\left[{\rm
      d}\Omega_\omega/{\rm d}\ln(k_{\rm phys})\right]$ plotted as a
      function of the rescaled variable $\rho $ characterizing the
      initial data. This plot corresponds to an inflationary era with
      a de Sitter metric, and Hubble parameter $\sim 10^{-6}M_{{\rm
      Pl}}$. The other parameter of initial data is $\Phi=\Phi _{\rm
      min}$. Allowed regions correspond to $
      \log_{10}\left(\vert\beta_k\vert^2 / \vert\beta_k^{\rm
      max}\vert^2\right)<0$, and are peaked around a particular value
      of $\rho $. The minimum is in fact $\vert\beta_k\vert^2=0$,
      corresponding to $\log_{10}\left(\vert\beta_k\vert^2 /
      \vert\beta_k^{\rm max}\vert^2\right)=-\infty$, but cannot be
      seen in the figure due to insufficient resolution. The dotted
      line provides a continuation of the numerical result to the
      analytical value at that point. In most of parameter space, the
      energy density is too large by $\sim$10 orders of
      magnitude. Right panel: Same as left panel for $H_0\sim
      10^{-61}M_{{\rm Pl}}$ in a matter dominated era. In nearly all
      of parameter space the energy density is too large by $\sim 122$
      orders of magnitude \label{fig:f1_dS}.}
\end{figure}

\begin{figure}
      \centering
      \includegraphics[width=0.49\textwidth,clip=true]{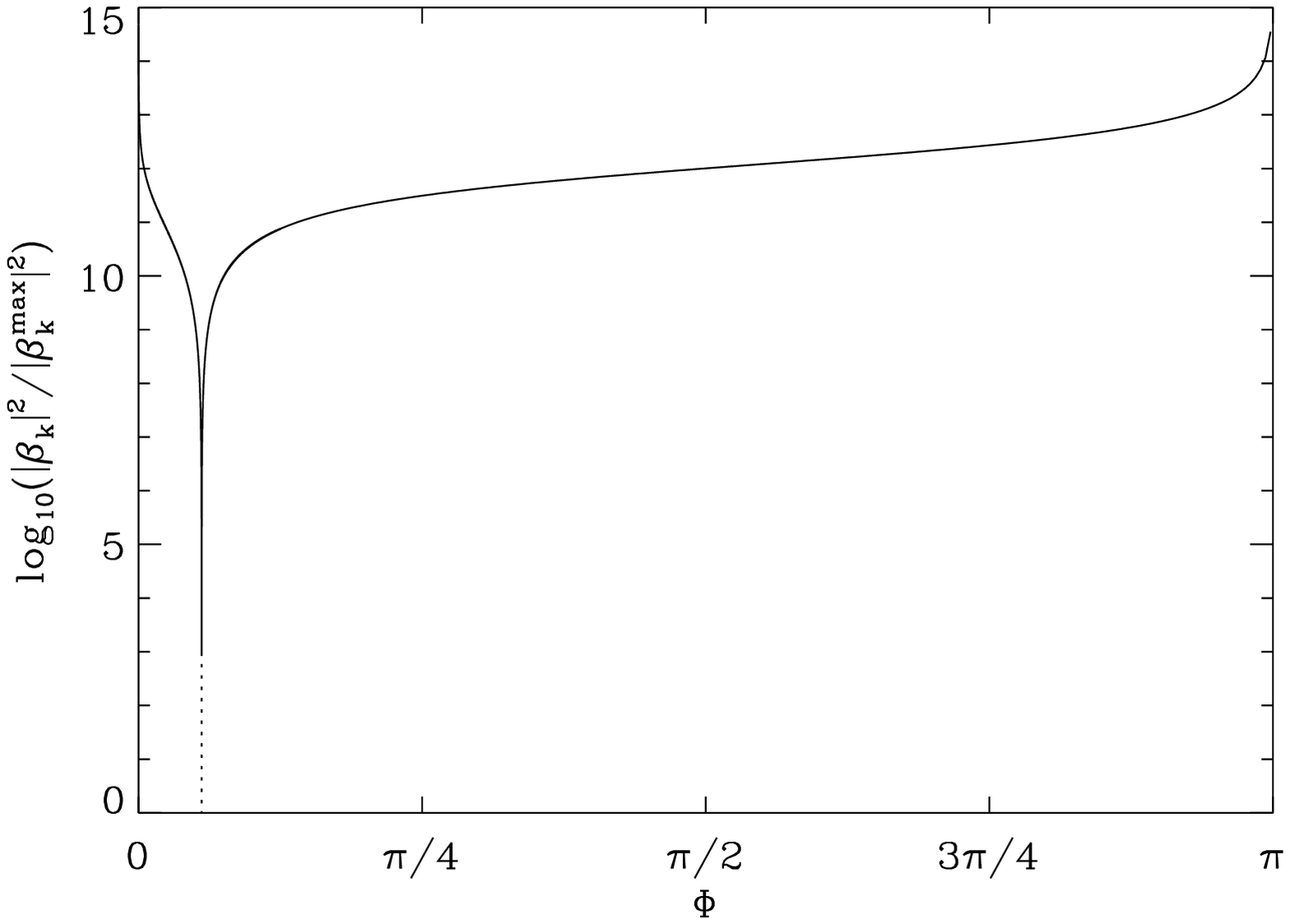}
      \includegraphics[width=0.49\textwidth,clip=true]{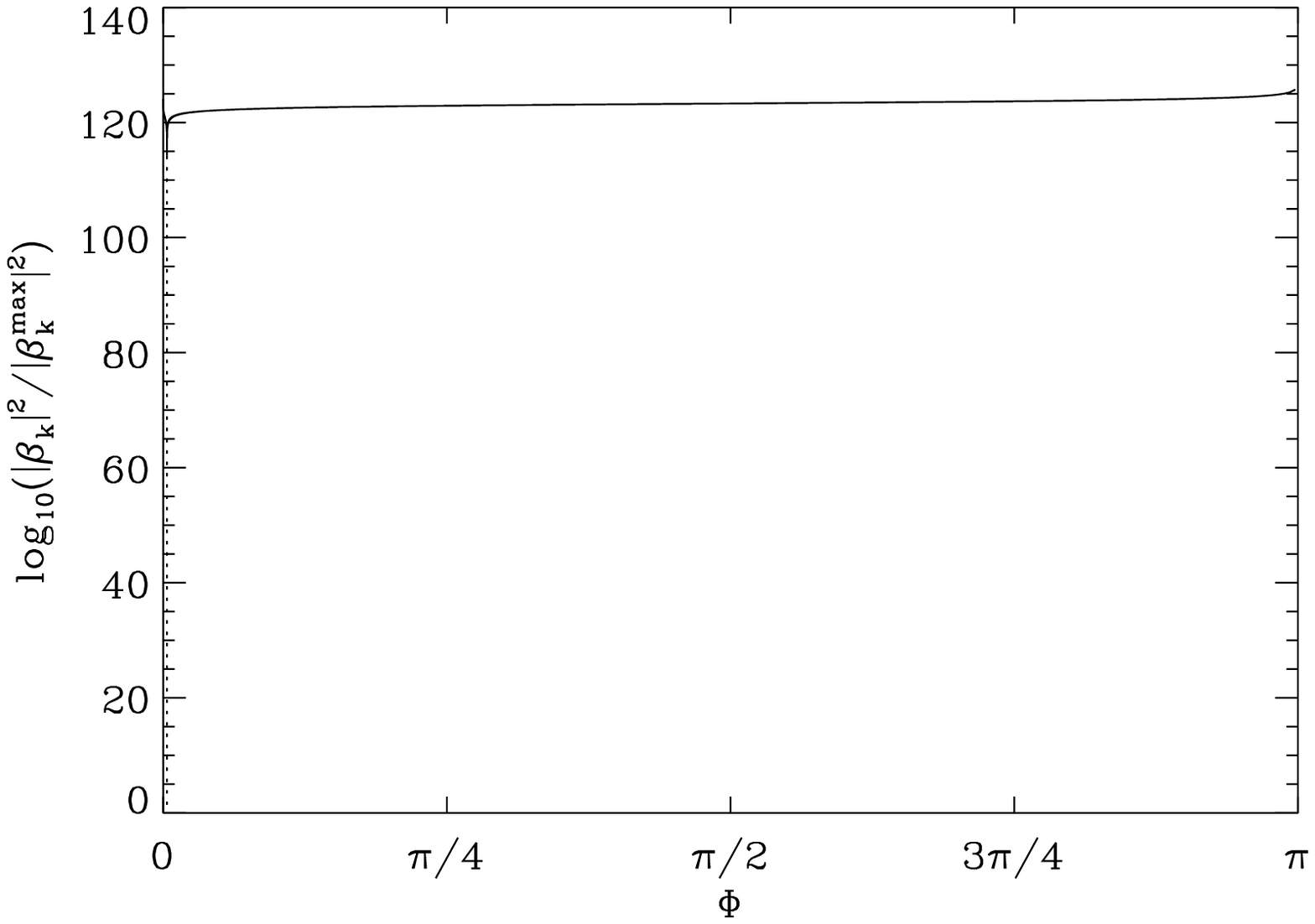}
      \caption[...]{Left panel: $\log_{10}\left(\vert\beta_k\vert^2 /
      \vert\beta_k^{\rm max}\vert^2\right)\equiv \left[{\rm
      d}\Omega_\omega/{\rm d}\ln(k_{\rm phys})\right]$ plotted as a
      function of the phase characterizing the initial data. This plot
      corresponds to an inflationary era with a de Sitter metric, and
      Hubble parameter $\sim 10^{-6}M_{{\rm Pl}}$. As indicated in the
      text, the other parameter of initial data (a modulus) has been
      eliminated for the phase $\Phi$ by minimizing
      $\vert\beta_k\vert^2$. Allowed regions correspond to $
      \log_{10}\left(\vert\beta_k\vert^2 / \vert\beta_k^{\rm
      max}\vert^2\right)<0$, and are seen to be peaked around a
      particular value of $\Phi$. The plot is symmetric in the
      interchange $\Phi\leftrightarrow \Phi+\pi$. Note that the
      ordinate scale is in $\log_{10}$: in most of parameter space,
      the energy density is too large by $\sim$10 orders of
      magnitude. Right panel: Same as left panel for $H_0\sim
      10^{-61}M_{{\rm Pl}}$ in a matter dominated era. In nearly all
      of parameter space the energy density is too large by $\sim 122$
      orders of magnitude, i.e. ${\rm d}\Omega_\omega/{\rm
      d}\ln(k_{\rm phys})\sim M_{{\rm
      Pl}}^4$\label{fig:f1_minimal_dS}.}
\end{figure}
\hspace{-5.05mm} and plot ${\rm d}\Omega_\omega/{\rm d}\ln(k_{\rm
phys})$ for $\vert\beta_k[\rho,\Phi _{\rm min}(\rho )]\vert^2$ in
Figs.~\ref{fig:f1_dS} and $\vert\beta_k[\rho_{\rm
min}(\Phi),\Phi]\vert^2$ in Figs.~\ref{fig:f1_minimal_dS} (for
respectively de Sitter inflation and today). One clearly sees from
these figures that ${\rm d}\Omega_\omega/{\rm d}\ln(k_{\rm phys})\sim
M_{{\rm Pl}}^2/H^2$ for most values of $\Phi$, which corresponds to
our previous expectations, i.e. the amount of energy density created
in quanta with $k_{\rm phys}\sim k_{\rm c}\sim M_{{\rm Pl}}$ is of
order $M_{{\rm Pl}}^4$. The behavior of $\vert\beta_k\vert^2$ around
the local minimum can be studied in the same way as before and one
obtains:
\begin{eqnarray}
\vert\beta_k\vert^2 (\rho ) &\simeq & 
\frac{1}{\beta ^2}\biggl(\frac{\beta }{1+\beta }\biggr )^{1/2}
(\rho -\rho _{\rm min})^2\, , 
\quad\quad (\rho \sim \rho _{\rm min})\, ,
\\
\vert\beta_k\vert^2 (\Phi ) &\simeq &
\frac14 \biggl(\frac{\beta}{1+\beta}\biggr)^2
\left[\left(1+\frac{K_+}{4k_{\rm c}}\right)^2
+ \frac{1+\beta}{\beta}\right]^2\left(\Phi-\Phi_{\rm
min}\right)^2\, , \quad\quad (\Phi\sim \Phi_{\rm min}).
\end{eqnarray}
If we write the value $\vert\beta_k\vert^2_{\rm max}$ such that ${\rm
d}\Omega_{\omega}/{\rm d}\ln(k_{\rm phys})=1$ (i.e. similar amount of
energy created in quanta with $k_{\rm phys}\sim k_{\rm c}$ than in the
background), then this value is reached if $\rho $ and $\Phi$ departs
from $\rho _{\rm min}$ and $\Phi_{\rm min}$ by an amount $\delta \rho
$, $\delta\Phi$, with:
\begin{equation}
\delta\rho \sim {\cal O}\left(\frac{H}{M_{{\rm
Pl}}}\right)\, ,
\quad 
\delta\Phi \sim {\cal O}\left[\frac{H}{M_{{\rm
Pl}}}\ln^{-2}\left(\frac{M_{{\rm Pl}}}{H}\right)\right].
\end{equation}
In other words, the fine-tuning necessary to avoid back-reaction is of
order $\sim \delta \Phi\sim \delta \rho \sim H/M_{{\rm Pl}}$
(neglecting the ln term). During inflation $H/M_{{\rm Pl}}\sim
10^{-6}$, and today $H/M_{{\rm Pl}}\sim 10^{-61}$. Although the
fine-tuning can be considered as not too severe during inflation, we
see that, during the matter epoch, it is as severe as the usual
fine-tuning of the cosmological constant problem. Therefore, the
scenario of Refs.~\cite{MBK01,BFM02,BM02} does not improve the
situation.

\par

At this point, it should be emphasized that the above problem is a
generic feature of the dispersion relation used in
Refs.~\cite{MBK01,BFM02,BM02}. In order for the trans-Planckian
effects to modify the power spectrum of the fluctuations, the physical
modes of interest must spend some time in a region where the WKB
approximation is violated. As already mentioned, this implies
production of particles and the energy density associated to these
particles must not exceed the background energy density. This implies
some constraints on the occupation number $\vert \beta _k\vert ^2$.
It has been shown in Ref.~\cite{S01} that these constraints are quite
stringent if the production is taking place today. Usually, these
tight constraints can be avoided by requiring that the violation only
occurs during inflation where the problem is less severe. This can be
achieved if the dispersion relation is such that $\omega_{\rm phys}\gg
H$ for all trans-Planckian wave-numbers, with $H$ the Hubble constant
some unspecified time after inflation. Such an example is provided in
Fig.~2 of Ref.~\cite{LMMU01}. Then as was argued in the discussion
following Eq.~(\ref{eq:omega-approx}) the WKB approximation should be
valid at all times after inflation and adiabaticity restored. Here
however, since $\omega _{\rm phys}\rightarrow 0$ as $k_{\rm
phys}\rightarrow +\infty$, there is at all times a region where the
WKB is violated. Therefore the class of dispersion relations used in
Refs.~\cite{MBK01,BFM02,BM02} suffer in a generic way from the problem
discussed in Ref.~\cite{S01}.

\par

Furthermore, we also note as previously that the above calculation of
the fine-tuning holds for a given co-moving wave-number $k$. Similar
constraints apply for other wave-numbers but the values of $\Phi_{\rm
min}$ and $\rho_{\rm min}$ are shifted from the above [notably because
of the choice $\eta_{\rm i}=\eta_{\rm m}(k)$]. Therefore one must not
only pick one right value of $\Phi$ to one part in $\sim M_{{\rm
Pl}}/H$, but a whole continuum of values of $\Phi(k)$ such that
back-reaction is avoided for all of these values.

\section{Equation of state}\label{Sec:IV}

Up to now we have argued that: {\it (i)} in the scenario proposed in
Refs.~\cite{MBK01,BFM02,BM02} there is no preferred initial vacuum
state, hence all conclusions drawn depend on the ad-hoc choice of
initial data; {\it (ii)} for arbitrary values of the two parameters
characterizing this choice of initial data one finds that energy in
excess of the background energy density is produced due to the non
adiabatic evolution of modes in the tail.

In a separate publication, we have constructed an effective
energy-momentum tensor for scalar field theories with non-linear
dispersion relations, and we have shown that dispersion relations of
the form of that proposed in Refs.~\cite{MBK01,BFM02,BM02} generically
led to the wrong equation of state. This finding has been challenged
by Bastero-Gil and Mersini recently, who argue that the
energy-momentum tensor we have constructed is ill-defined as it
(supposedly) is not divergenceless.

Explicitly, in Ref.~\cite{LMMU01}, the {\it vev} for the energy
density and pressure are given by
\begin{eqnarray}
 \langle\rho\rangle &=& \frac{1}{4\pi^2a^4}\int {\rm d} k k^2
     \left[a^2\left|\left(\frac{\mu_k}{a}\right)'\right|^2
     +\omega^2(k)\left|\mu_k\right|^2\right], \label{rhoeff}\\
 \langle p\rangle&=&\frac{1}{4\pi^2a^4}\int {\rm d} k k^2
      \left[a^2\left|\left(\frac{\mu_k}{a}\right)'\right|^2 +\left(
      \frac{2}{3}k^2\frac{{\rm d}\omega^2}{{\rm d}k^2}-\omega^2\right)
      \left|\mu_k\right|^2\right], \label{peff}
\end{eqnarray}
and the integrals extend from $0$ to $+\infty$. Reference~\cite{BM02}
claims that $\langle\rho\rangle '+ 3{\cal
H}\langle\rho+p\rangle\neq0$, with ${\cal H}\equiv a'/a$.  However,
noting that the co-moving frequency $\omega(k)= a \omega_{\rm
phys}(k_{\rm phys})$, a straightforward calculation gives:

\begin{eqnarray}
\langle\rho \rangle '& = & \frac{1}{4\pi^2a^4}\int {\rm d} k k^2
\biggl\{ \left[\mu_k'' - {\cal H}\mu_k' + \left({\cal H}^2 - {a''\over
a}\right)\mu_k\right]\left(\mu_k^{*\prime} - {\cal H}\mu_k^*\right) +
\left(\mu_k^{\prime} - {\cal H}\mu_k\right)
\left[\mu_k^{*\prime\prime} - {\cal H}\mu_k^{*\prime} + \left({\cal
H}^2 - {a''\over a}\right)\mu_k^*\right] \nonumber\\ & & - 4{\cal
H}\left\vert\mu_k' - {\cal H}\mu_k\right\vert^2 +
\omega^2(k)\left(\mu_k'\mu_k^*+ \mu_k\mu_k^{*\prime}\right) - 2{\cal
H}k^2 {{\rm d}\omega^2 \over {\rm d}k^2}\vert\mu_k\vert^2 -2{\cal
H}\omega^2(k)\vert\mu_k\vert^2\biggr\} \nonumber\\ & = &
\frac{1}{4\pi^2a^4}\int {\rm d} k k^2 \biggl\{ \left(\mu_k'' -
{a''\over a}\mu_k\right)\left(\mu_k^{*\prime} - {\cal H}\mu_k^*\right)
+ \left(\mu_k^{\prime} - {\cal H}\mu_k\right)
\left(\mu_k^{*\prime\prime} - {a''\over a}\mu_k^*\right) -6{\cal
H}\left\vert\mu_k' - {\cal H}\mu_k\right\vert^2 \nonumber \\ & &
+\omega^2(k)\left[\mu_k\left(\mu_k^{*\prime} - {\cal H}\mu_k^*\right)
+ \mu_k^*\left(\mu_k^\prime - {\cal H}\mu_k\right)\right] - 2{\cal
H}k^2 {{\rm d}\omega^2 \over {\rm d}k^2}\vert\mu_k\vert^2\biggr\}
\nonumber\\ & = & \frac{1}{4\pi^2a^4}\int {\rm d} k k^2 \biggl(
-6{\cal H}\left\vert\mu_k' - {\cal H}\mu_k\right\vert^2 - 2{\cal H}k^2
{{\rm d}\omega^2 \over {\rm d}k^2}\vert\mu_k\vert^2\biggr),
\end{eqnarray}
where the field equation $\mu_k''-(a''/a)\mu_k=-\omega^2(k)\mu_k$
has been used in the last step. Since
\begin{equation}
3{\cal H}\langle\rho + p\rangle\,=\, \frac{1}{4\pi^2a^4}\int {\rm
d} k k^2 \left( 6{\cal H}\left\vert\mu_k' - {\cal
H}\mu_k\right\vert^2 + 2{\cal H}k^2 {{\rm d}\omega^2 \over {\rm
d}k^2}\vert\mu_k\vert^2\right)\, ,
\end{equation}
the energy conservation condition $\langle\rho\rangle' + 3{\cal
H}\langle\rho+p\rangle=0$ is trivially satisfied.

We take advantage of this Section to point out that the
energy-momentum tensor we have constructed in a previous
publication~\cite{LMMU01} is well behaved and the construction is
entirely consistent. In effect we constructed a generally covariant
Lagrangian for a scalar field with a non-linear dispersion
relation. The breaking of the Lorentz invariance is explicited by
introducing a dynamical four-vector $u^\mu$ in the Lagrangian whose
role is to define the preferred rest frame while preserving general
covariance at the same time, following previous work by Jacobson and
Mattingly~\cite{JM01} (see also references therein). The
energy-momentum tensor derived by varying the action with respect to
the metric is the sum of the energy-momentum tensor of the scalar
field and that of $u^\mu$. We have restricted ourselves to FLRW
space-times by choosing $u^\mu=(-1,0,0,0)$. This approach is
consistent as $u^\mu$ satisfies its field equation [see Eq.~(15) in
Ref.~\cite{LMMU01}], and the scalar field also satisfies its field
equation. In FRLW space-time, the energy-momentum tensor of the
four-vector (i.e. the part which depends only on $u^\mu$) vanishes
since $u^\mu$ is constant~\cite{LMMU01}, hence the remainder is the
scalar field energy-momentum tensor, which is conserved as shown
above. We consider this scalar field as a test field, which we
quantize on the curved background. The dynamics of the background can
be specified by adding matter fields to the action without modifying
our approach. Indeed the matter field energy-momentum tensor will be
separately conserved. Therefore our earlier criticisms on the equation
of state of the trans-Planckian modes apply.

We also note that a recent paper by Frampton~\cite{F02} argues that by
adding higher order terms of the form $u^\mu{\cal D}^{2n}u_\mu$ in the
effective Lagrangian, one can obtain a correct equation of state,
i.e. similar to that of vacuum. Here ${\cal D}^2$ denotes a
three-dimensional Laplacian expressed in a generally covariant way on
space-like hyper-surfaces (see Refs.~\cite{JM01,LMMU01} for explicit
definitions). The choice $u^\mu=(-1,0,0,0)$ then reduces this term to
the usual three-dimensional Laplacian on constant time hyper-surfaces
for FLRW space-times. However, as should be obvious, the only terms
that can enter the energy-momentum tensor {\it via} this new term
always contain derivatives of the form ${\cal D}^{2n} u^\mu$ when $n\geq
2$. Those terms vanish when one picks $u^\mu$ to be constant as
above. For $n=1$, i.e. the lowest order term of the expansion, one can
check that the variation of ${\cal D}^2 u^\mu$ with respect to the
metric or its first derivative always induces a term proportional to a
derivative of $u^\mu$ that is either spatial or temporal. This
calculation can be found in Eqs.~(B3) and (B4) of Ref.~\cite{LMMU01}
(the substitution $\phi\rightarrow u^\mu$ in this calculation can be
made without changing the result since the equation is written in a
generally covariant way). Therefore we conclude that the addition of
terms proposed by Frampton cannot lead to any difference with respect
to the energy-momentum tensor we proposed earlier. Hence such terms
cannot account straightforwardly for a vacuum-like equation of state in
this scenario.

The difference with the conclusion of Ref.~\cite{F02} probably lies in
the choice of normalization of $u^\mu$: Ref.~\cite{F02} claims that,
in a cosmological context, $u^\mu u_\mu - a^{-2} = 0$ ($a$ is the
scale factor). However, $u^\mu u_\mu$ is a scalar, while $a^{-2}$ is
the component of a rank two tensor so that this choice of
normalization is manifestly not covariant. Of course, it can be made
covariant at the expense of the introduction of a second vector field
but this does not seem to be the case in Ref.~\cite{F02}. This implies
that the Lagrangian proposed by Frampton is generally not
covariant. Of course the correct normalization for $u^\mu$ in a
cosmological context is always $u^\mu u_\mu - 1 =0$ (up to the sign
convention), and $u^\mu$ can be written as $(-1/a,0,0,0)$ if the FRLW
metric is written in terms of conformal time, or $(-1,0,0,0)$ if the
metric is written in terms of cosmic time.

\par
 
To finish, let us also note that it is claimed in Ref.~\cite{MBK01}
that ``{\it the tail modes are still frozen at present... Thus the
energy of the tail is a contribution to the dark energy of the
Universe: up to the present it has the equation of state of a
cosmological constant term}''.  Here, that the tail modes are frozen
means $(\mu_k/a)'\sim 0$, the solution to the field equation when the
term $a''/a$ dominates (for $\xi=0$). But there is no logical
relationship between these modes being frozen and the equation of
state being that of a cosmological constant. As a matter of fact the
equation of state of the modes of wavelengths larger than the horizon
size for a scalar field with a linear dispersion relation takes the
form $p=-\rho/3$. The derivation of the equation of state requires to
define correctly a stress-energy tensor when such non-linear
dispersion relations are taken into account (and thus when Lorentz
invariance is broken) as we did in Ref.~\cite{LMMU01}. Note also that
there is a contradiction between the above claim of Ref.~\cite{MBK01}
and the whole content of the recent paper by Bastero-Gil and
Mersini~\cite{BM02} which calculates the equation of state of the
trans-Planckian modes and finds that it is not the equation of state
of the vacuum but that it approaches it only at late times.

\section{Conclusions}
\label{Sec:V}

In this article, we have examined in detail the scenario proposed in
Refs.~\cite{MBK01,BFM02,BM02} which attributes the dark energy to the
properties of a scalar field with a dispersion relation that decreases
exponentially with trans-Planckian wave-number $k_{\rm phys}\gtrsim
k_{\rm c}\sim M_{\rm Pl}$. We have demonstrated that this mechanism
does not work, mainly for two reasons. (i) The mode function of the
scalar field does not behave as a plane wave as
$\eta\rightarrow-\infty$ in the so-called ``tail'' (i.e., the part of
the non-linear dispersion relation where $\omega_{\rm phys}\ll H$ and
$k_{\rm phys}\gg M_{{\rm Pl}}$). This implies that there is no
definite prescription for constructing a well-defined initial vacuum
state, hence that the choice of initial data is entirely arbitrary.
This situation is similar to the problem of setting initial data for
cosmological perturbations in the absence of an accelerated expansion
era (but with linear dispersion relations), in which case the data
would have to be specified on super-horizon scales where the mode
function does not oscillate and is frozen by the expansion. In the
scenario of Refs.~\cite{MBK01,BFM02,BM02} the WKB approximation is not
valid at all times for modes in the tail. This explains that the
notion of adiabatic vacuum cannot be used to set up initial data and
the initial state proposed in Ref.~\cite{MBK01} is thus ad-hoc. It
also brings us to the second objection against this scenario: (ii)
since all modes originate in the tail of the dispersion relation, the
breakdown of the WKB approximation for a given physical wave-number at
early times implies the continuous production of a substantial number
of quanta with physical wave-numbers $\gtrsim k_{\rm c}$. The
breakdown of the WKB approximation can indeed be seen as the signal of
a strongly non-adiabatic evolution which is generically associated
with particle production in expanding space-times. We have calculated
the amount of energy density produced in modes of physical wave-number
$k_{\rm c}$ as a function of the two parameters that characterize the
(arbitrary) choice of initial data. We have shown that this energy
density is generically of order $M_{{\rm Pl}}^4$.  The production of
energy density in quanta with wave-numbers $\sim M_{{\rm Pl}}$ in
excess of the background energy density $\sim H^2 M_{{\rm Pl}}^2$
implies the breakdown of the perturbative semi-classical framework
used for the calculation, and renders all claims irrelevant.  There is
a small region of parameter space in which this energy density can be
tuned down to zero, but the fine-tuning in the choice of initial data
is of order $H/M_{{\rm Pl}}$. Such a fine-tuning at the time of
inflation is of order $\sim 10^{-6}$ and is probably acceptable, but
today, it is of the same order as the celebrated fine-tuning of the
cosmological constant problem.

\end{document}